\documentclass[lettersize,journal]{IEEEtran}
\usepackage{amsmath,amsfonts}
\usepackage{algorithmic}
\usepackage{array}
\usepackage{textcomp}
\usepackage{stfloats}
\usepackage{url}
\usepackage{verbatim}
\usepackage{graphicx}
\usepackage{cite}
\usepackage{xcolor}

\def\BibTeX{{\rm B\kern-.05em{\sc i\kern-.025em b}\kern-.08em
    T\kern-.1667em\lower.7ex\hbox{E}\kern-.125emX}}
\usepackage{balance}

\usepackage{siunitx}
\usepackage{longtable,tabularx}

\usepackage{subcaption}
\usepackage{soul}

\usepackage{hhline} 
\usepackage{multirow} 

\usepackage{amsthm}

\theoremstyle{definition}
\newtheorem{thm}{Theorem}
\newtheorem{lem}[thm]{Lemma}

\newtheorem{rem}{Remark}
\newtheorem{assumption}{Assumption}

\usepackage{siunitx}
\usepackage{cleveref}
\usepackage{bm}

\newcommand{\graph}{\mathcal{G}}
\newcommand{\vertex}{\mathcal{V}}
\newcommand{\edge}{\mathcal{E}}
\newcommand{\laplacian}{\mathcal{L}}
\newcommand{\N}{\mathcal{N}}
\renewcommand{\vec}{\mathbf}

\begin{document}
\title{Cooperative Salvo Guidance over Leader-Follower Network with Free-Will Arbitrary Time Convergence}
\author{Rajib Shekhar Pal\thanks{Research Scholar, Intelligent Systems and Control Lab, Department of Aerospace Engineering, \textbf{email}: \texttt{rajibspal@iitb.ac.in}.},  Shashi Ranjan Kumar\thanks{Associate Professor, Intelligent Systems and Control Lab, Department of Aerospace Engineering, \textbf{email}: \texttt{srk@aero.iitb.ac.in}.}, and Dwaipayan Mukherjee\thanks{Assistant Professor, Department of Electrical Engineering, \textbf{email}: \texttt{dm@ee.iitb.ac.in}.} \\
{\normalsize\itshape Indian Institute of Technology Bombay, Powai-- 400076, Mumbai, India}}


\maketitle

\begin{abstract}
	A cooperative salvo strategy is proposed in this paper which achieves consensus among the interceptors within a pre-defined arbitrary settling time. Considering non-linear engagement kinematics and a system lag to capture the effect of interceptor autopilot as present in realistic interception scenarios, the guidance schemes use the time-to-go estimates of the interceptors in order to achieve simultaneous interception of a stationary target at a pre-determined impact time. The guidance scheme ensures that consensus among the time-to-go estimates of the interceptors is achieved within a settling time whose upper bound can be pre-specified arbitrarily independent of the initial conditions or design parameters. The efficacy of the proposed guidance strategy is demonstrated using numerical simulations with varied conditions of initial position, velocities and heading angle errors of the interceptors as well as different desired impact times.
\end{abstract}

\def\abstractname{Note to Practitioners}
\begin{abstract}
	The paper is motivated by the problem of simultaneous interception of a target by multiple interceptors, known as salvo attack. The synchronization of the interceptors' maneuvers in order to achieve such simultaneous interception within a desired time is an interesting research problem among the guidance community. The existing approaches are able to achieve such synchronization within a finite time. However, the upper bound on the finite time cannot be chosen freely, and is dependent either on both design parameters and initial conditions or on the design parameters alone. In this paper, we develop an algorithm that uses a leader-follower communication network in order to achieve synchronization in the estimated time-to-go of the interceptors, i.e. the time remaining till interception of the target. The time of synchronization and the desired interception time can be predetermined by the designers at the onset of the maneuvers. Preliminary simulations show that the algorithm is able to achieve interception of a stationary target under various initial conditions as well as considering first-order autopilot dynamics of the interceptors. Future research will address salvo attack with maneuvering targets considering bounded control. 
\end{abstract}


\section{Introduction}
\label{sec:introduction}

\IEEEPARstart{C}{ooperation} during hunting is observed in nature among various predators as it proves to be a more efficacious strategy than attacking alone against a larger and stronger prey. Simultaneous interception of the target, known as a \emph{salvo attack}, can similarly overwhelm and exhaust a stronger target's defense capabilities and counter its evasive tactics. This idea has led to a growing interest in the salvo guidance problem among researchers within the guidance community \cite{Zhang2015, Song2015, Zhu2017, He2018a, He2018, Ai2019, Jeon2010, Hou2015, Li2016, Kumar2020, Li2018a}. 

The impact-time controlled guidance laws designed for target interception by a single interceptor, achieved using the widely popular sliding mode control \cite{Kumar2015, Kim2019, Chen2018, Hu2019}, optimal control \cite{Jeon2006, Liu2017b} and adaptive control \cite{Lu2006} algorithms, can be modified to achieve salvo attacks by ensuring the impact time values of individual interceptors are identical. However, in the event of any interceptor deviating from its desired course, for any reason, no corrective action is initiated by the other interceptors to still ensure salvo. This may lead to the failure of the salvo mission. Thus, simultaneous attack or salvo in non-ideal conditions, using these schemes, may not be guaranteed since there is no communication or coordination among the interceptors. In \cite{Zhang2015}, the authors used biased proportional navigation (PN) to develop a cooperative guidance scheme, while the problem of impact time control was posed as a range tracking problem in \cite{Zhu2017} for the achievement of salvo attack. A cooperative time-invariant guidance scheme was developed in \cite{Song2015} using time-to-go-estimates. In order to avoid using the time-to-go estimate explicitly in designing cooperative salvo attack strategies, a two-stage guidance approach was used in \cite{He2018a, He2018}. A decentralized control law was used in the first stage for achieving salvo and this stage provided suitable initial conditions for the second stage. Classical PN guidance law was then used in the second stage to achieve simultaneous target interception. The authors in \cite{Ai2019} proposed a similar two-stage salvo attack scheme using optimal control, with field-of-view constraint in the first stage and classical PN guidance in the second stage. In \cite{Yang2023}, a multi-agent consensus-based algorithm was used in developing a cooperative guidance scheme with impact-angle and field-of-view constraints. This guidance scheme also used a two-stage strategy where consensus was achieved in time-to-go and lead angle in the two stages, respectively. However, the guidance strategies proposed in \cite{Zhang2015, Song2015, Zhu2017, He2018a, He2018, Ai2019, Yang2023} could only achieve asymptotic convergence of the relevant error variables.

Finite-time stability is a stronger notion than asymptotic stability, with better convergence guarantees and disturbance rejection properties \cite{Bhat2000}. Guidance law for a single interceptor with finite time convergent line-of-sight angular rate was developed in \cite{Zhou2009}. For salvo attack, finite-time consensus-based cooperative guidance schemes were developed for both directed and undirected communication topologies. Authors in \cite{Jeon2010} proposed a cooperative PN guidance strategy with time-varying navigation gain which minimizes time-to-go variance over an undirected network. This was modified in \cite{Hou2015} to develop a salvo guidance algorithm that achieves finite-time convergence, constrained by acceleration saturation. A cooperative salvo guidance law was presented in \cite{Kumar2020} over a directed cycle graph to achieve consensus in time-to-go and intercept the target at a desired finite time, by appropriately selecting the edge weights of the graph.

One drawback of finite-time convergent control systems is that the upper bound on the settling time is dependent on the initial conditions and design parameters. The dependence on initial conditions was eliminated by fixed time stable control laws \cite{Polyakov2012}, and this concept has since gained prominence in cooperative control of multi-agent systems \cite{Zuo2018} and cooperative salvo guidance, in particular. The authors in \cite{Li2018a} proposed an adaptive fault-tolerant cooperative guidance scheme for simultaneous target interception that was able to achieve fixed time convergence under partial actuator failure. In \cite{Wang2022a}, a distributed impact-angle-constrained salvo attack strategy was developed by augmenting the optimal impact angle control guidance law with a cooperative guidance term which is able to ensure that the consensus of time-to-go among the multiple interceptors occurs within a fixed time prior to interception. However, in the fixed time convergence paradigm, the upper bound on settling time cannot be selected arbitrarily as it still remains a function of design parameters.

The concept of fixed time stability has recently been extended to free-will arbitrary time convergence in \cite{Pal2020a} which allows the designer to choose the upper bound on settling time arbitrarily, irrespective of initial conditions or design parameters. This technique was used in \cite{Tran2022} and \cite{Pal2022} to develop consensus algorithms for multi-agent systems with single and double integrator dynamics. Their algorithms were able to achieve consensus in the agents' states within a fixed time that has an arbitrarily pre-specified upper bound. However, the algorithms were developed over undirected communication topologies among the multiple agents.

In this paper, a cooperative guidance algorithm is proposed for salvo attack on a stationary target, based on time-to-go estimates. This achieves consensus in time-to-go among multiple interceptors within a free-will arbitrarily specified time, independent of initial conditions or design parameters. Thus, target interception occurs precisely at a desired pre-specified impact time. The interceptors share information among each other in a leader-follower communication topology. The leader sends its state information to one or more followers unidirectionally, while the followers communicate among each other over an undirected connected graph. Two types of interceptor dynamics are considered for designing the guidance schemes -- one with an ideal autopilot, and the other with first-order autopilot dynamics. The latter set-up allows us to account for approximate autopilot lag present in practical scenarios. Extensive numerical simulations with various initial conditions and system parameters are performed to establish the efficacy of the proposed guidance strategies. Some preliminary results were reported in \cite{pal2023acc}.

The contributions of the paper can be summarized as follows.
\begin{enumerate}
    \item The existing cooperative salvo algorithms can only achieve finite time or fixed time convergence in time-to-go consensus. The proposed algorithm achieves consensus within a settling time than can be pre-specified arbitrarily by the user and is independent of initial conditions as well as any design parameters.

    \item The existing literature on free-will arbitrary time consensus uses the assumption of undirected communication among the agents. These algorithms can only achieve average consensus in time-to-go and would not allow its convergence on any desired value. The proposed leader-follower consensus algorithm solves this problem and makes possible simultaneous interception of the target at the desired time precisely.

    \item The existing salvo guidance algorithms require the desired impact time to lie within the range of initial times-to-go of the interceptors. The proposed algorithm is not limited by such condition and can achieve target interception at desired impact time outside the bounds.

    \item The proposed guidance strategy considers realistic interceptor models with first-order autopilot dynamics.

    \item Starting with a wide range of positions and heading angles, interception of a stationary target is guaranteed.

    \item Efficacy and scalability of the proposed algorithm is demonstrated using large number of interceptors.

\end{enumerate}

\section{Preliminaries}
\label{sec:preliminaries}

The communication topology among the interceptors is described by a graph, $\graph$, consisting of the vertex set, $\vertex$, and the edge set, $\edge \subseteq \vertex \times \vertex$. The edge $(i,j)$ indicates that the $j^{\rm th}$ interceptor receives state information from the $i^{\rm th}$ interceptor which is then termed as the neighbor of the $j^{\rm th}$ interceptor. Bidirectional communication among the follower interceptors means the edge $(i,j) \in \edge \iff (j,i) \in \edge$. $\N_i=\{ j \in \vertex : (i,j) \in \edge \}$ is the set of neighbors of the $i^{\rm th}$ interceptor. The interaction among the interceptors is algebraically encapsulated in the interaction matrix $\bm{\mathcal{H}}$ which consists of two parts given by $\bm{\mathcal{H}} = \bm{\mathcal{L}_{f}} + \bm{\mathcal{B}}$. Here, $\bm{\mathcal{L}_{f}}$ is the graph Laplacian matrix \cite{Godsil2001} associated with the connected, undirected, and unweighted graph among the follower interceptors, while $\bm{\mathcal{B}}$ denotes the leader incidence matrix depicting flow of information from the leader to a subset of the follower interceptors. The Laplacian matrix $\bm{\mathcal{L}_{f}}$ and the leader incidence matrix are defined as
\begin{align*}
[\bm{\laplacian_f}]_{ij} &= \begin{cases}
-1, & \text{if}~ (i,j)\in \edge\\
-\sum_{j \in \N_i} [\bm{\laplacian_f}]_{ij}, &\forall~i=j \\
0, &\text{otherwise}
\end{cases},  
\\
B_{ij} &= \begin{cases}
1, & \text{if } i=j \text{ and directed edge from leader to $i^{\rm th}$ follower} \\
0, & \text{ otherwise }
\end{cases}
\end{align*}
The undirected graph is said to be connected if a \emph{path} (i.e., an alternating sequence of nodes and edges such that the edge between two nodes in the sequence is the edge connecting the nodes) exists between any two vertices of the graph \cite{Godsil2001}. For such a graph, the Laplacian matrix is symmetric, positive semi-definite with real eigenvalues given by $0 = \lambda_1 < \lambda_2 \le \lambda_3 \le \ldots \le \lambda_n$. The eigenvector corresponding to $\lambda_1 = 0$ of $\bm{\laplacian_f}$ is $\mathbf{1}_n$ such that $\bm{\laplacian_f} \mathbf{1}_n = \mathbf{0}_n$. However, the leader-follower interaction matrix $\bm{\mathcal{H}}$ is a symmetric \emph{positive definite} matrix, with real, positive eigenvalues.

The following lemmas will aid in the synthesis of our guidance strategies.
\begin{lem}[\cite{Pal2020a} Theorem 1]
\label{lemma:fwat_convergence}
Consider the nonlinear dynamical system $\dot{\mathbf{x}} = \mathbf{f}(t,\mathbf{x},\bm{\alpha}),~\mathbf{x}(t_0)=\mathbf{x}_0 $ where $\mathbf{x} \in \mathbb{R}^n$ denotes the system states, and $f:\mathbb{R}_{\ge 0} \times \mathbb{R}^n \to \mathbb{R}^n$ and $\bm{\alpha} \in \mathbb{R}^l$ denote the adjustable system parameters. Let $\mathbf{x}=\mathbf{0}$ be an equilibrium point of the system within the domain $\mathcal{D} \subseteq \mathbb{R}^n$. Let $\beta_1(\mathbf{x})$ and $\beta_2(\mathbf{x})$ be two continuous positive definite functions on $\mathcal{D}$ and $\eta \ge 1$. Suppose that there exists $V(t,\mathbf{x}):[t_0,t_f) \times \mathcal{D} \to \mathbb{R}_{\ge 0}$ which is a real-valued continuously differentiable function such that
\begin{align}
& \beta_1(\mathbf{x}) \le V(t,\mathbf{x}) \le \beta_2(\mathbf{x}), ~ \forall ~ t \in [t_0,t_f),\\ &V(t,\mathbf{0}) = 0, ~ \forall ~ t \in [t_0,t_f), \\
\label{eq:fwat_convergence_vdot_inequality}
& \dot{V}(t,\mathbf{x}) \le -\frac{\eta}{t_f - t}(1-e^{-V(t,\mathbf{x})}), 
\end{align}
$\forall \mathbf{x} \in \mathcal{D}, ~ \forall t \in [t_0,t_f)$ then $\mathbf{x}=\mathbf{0}$ is free-will arbitrary time stable with $t_f > t_0$ being an arbitrary pre-specified time. If the equality in \eqref{eq:fwat_convergence_vdot_inequality} holds strictly for all $\mathbf{x} \in \mathcal{D},~\forall t \in [t_0,t_f)$, then convergence to $\mathbf{x}=\mathbf{0}$ occurs at $t_f$.
\end{lem}
\begin{lem}[\cite{Tran2022} Lemma 2]
\label{lemma:single_var_inequality}
For any $x,y \in \mathbb{R}$ satisfying $0 < x \le y$, the following holds $-x(1-e^{-x}) \ge -y(1-e^{-y}).$
\end{lem}
\begin{lem}[\cite{Tran2022} Lemma 3]
\label{lemma:multi_var_inequality}
For any vector $\mathbf{x} \in \mathbb{R}^n$, the following holds, $-||\mathbf{x}||(1-e^{-||\mathbf{x}||})  \ge -\mathbf{x}^T(I_n - e^{-\text{diag}(\mathbf{x})})\mathbf{1}_n \ge -\mathbf{x}^T(\mathbf{1}_n - e^{-\mathbf{x}})$
where $e^{-\text{diag}(\mathbf{x})}\mathbf{1}_n = e^{-\mathbf{x}}\in\mathbb{R}^n$.
\end{lem}

\section{Problem Formulation}
\label{sec:problem_formulation}
\begin{figure}[t!]
	\centering
	\includegraphics[width=0.45\textwidth]{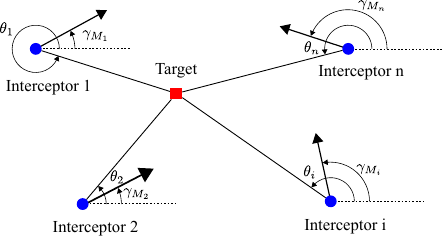}
	\caption{Planar engagement geometry.}
	\label{fig:engagement_geometry}
\end{figure}

We consider the planar engagement of interceptors against a stationary target in this paper as illustrated in Fig.~\ref{fig:engagement_geometry}. The $i^{\rm th}$ interceptor is at a radial distance of $r_i$ from the stationary target while moving at a constant speed of $V_{M_i}$. Its flight path angle is denoted by $\gamma_{M_i}$, while $\theta_i$ is the line-of-sight (LOS) angle with respect to a fixed reference. With $a_{M_i}$ as the lateral acceleration, the engagement kinematics of the $i^{\rm th}$ interceptor can be written as
\begin{align}
    \dot{r}_i &= V_{r_i} = -V_{M_i}\cos \theta_{M_i},\\
    r_i\dot{\theta}_{i} &= V_{\theta_i} = -V_{M_i}\sin\theta_{M_i}, \\
    \dot{\gamma}_{M_i} &= \frac{a_{M_i}}{V_{M_i}},
\end{align}
where $\theta_{M_i} = \gamma_{M_i} - \theta_i$ denotes the heading angle error. The terms $V_{r_i}$ and $V_{\theta_i}$ represent the parallel and perpendicular components of the relative velocity with respect to the LOS, respectively. The following assumptions are made.

\begin{assumption}
\label{assumption:point_mass_constant_speed}
The interceptors are point-mass vehicles with constant speeds throughout the engagement.
\end{assumption}

\begin{assumption}
\label{assumption:leader_global_info}
There is a single leader, while the rest of the interceptors are followers who communicate among themselves over an undirected, connected graph. The desired terminal time is known to the leader alone.
\end{assumption}

\begin{assumption}
\label{assumption:communication_flow}
Information flows from the leader to one or multiple followers in a unidirectional communication.
\end{assumption}

\begin{assumption}
\label{assumption:no_delay}
Communication among the interceptors occur without any delay or failure.
\end{assumption}

Time-to-go is the total time remaining until the interceptor intercepts the target and is denoted by $t_{go_i}$ \cite{Kumar2020}. The impact time, defined as $t_{imp_i} = t_{{\rm el}_i}+t_{go_i}$, is the total time taken by an interceptor since its launch till interception of the target,  \cite{Kumar2020}. Here, $t_{{\rm el}_i}$ denotes the elapsed time since the $i^{\rm th}$-interceptor's launch. With $T_d$ as the desired impact time, the desired time-to-go $t_{go_d}$ can be written as $t_{go_d} = T_d - t_{{\rm el}_i}$. Consequently, we have $\dot{t}_{go_d} = -1$ and $\ddot{t}_{go_d} = 0$.

The main problem addressed in this paper may now be stated. Subject to the assumptions \ref{assumption:point_mass_constant_speed}-\ref{assumption:communication_flow}, design a cooperative salvo guidance strategy for interception of a stationary target by multiple interceptors who interact with each other over a leader-follower communication topology, considering the following types of autopilot dynamics.
 \begin{enumerate}
    \item Each interceptor's autopilot and rotational dynamics are sufficiently fast   compared to the guidance command. Hence, the desired acceleration, $a_{M_i}$, is achieved instantaneously, and the autopilot dynamics is ideal.
    
    \item The following first-order model approximates the lumped rotational and autopilot dynamics of each interceptor as
    \begin{align}
    \label{eq:first_order_autopilot_dynamics}
    \dot{a}_{M_i} = \frac{a_{Mc_i} - a_{M_i}}{\tau_i},
    \end{align} 
    where $a_{Mc_i}$ and $a_{M_i}$ denote the commanded and achieved interceptor lateral accelerations, respectively, while $\tau_i$ is the autopilot time constant.
 \end{enumerate}

\section{Main Results}
\label{sec:main_results}
In this section, we present guidance designs to achieve guidance objectives while considering ideal and first-order autopilot dynamics. The actual value of the time-to-go can be back-calculated exactly only after the completion of the engagement. Therefore, an estimate is provided in \cite{Jeon2010} for the $i^{\rm th}$ interceptor as
\begin{align}
\label{eq:tgo_estimate}
t_{go_i} = \frac{r_i}{V_{M_i}}\left(1 + \frac{\theta_{M_i}^2}{2K}\right),~K = 2N-1,
\end{align}
where $N=3$ is the navigation constant for proportional navigation (PN) guidance. It may be noted here that \eqref{eq:tgo_estimate} uses the small angle approximation of $\theta_{M_i}$. However, it has been widely used in literature \cite{Hou2015,Li2018a,Kumar2020}, which shows that it works satisfactorily in the development of salvo guidance algorithms.

\subsection{Salvo guidance with ideal autopilot}

Differentiating \eqref{eq:tgo_estimate} with respect to time, we get
\begin{align}
\label{eq:tgodot_estimate}
\dot{t}_{go_i} &= F_i + B_i a_{M_i},\\
F_i &= \frac{V_{r_i}t_{go_i}}{r_i} - \frac{ V_{\theta_i} \theta_{M_i} }{ V_{M_i} K },~B_i = \frac{ r_i\theta_{M_i} }{ V_{M_i}^2 K}.
\end{align} 
Let $t_{go_{\ell}} \in \mathbb{R}$ denote the time-to-go for the leader interceptor. With the desired time-to-go as $t_{go_{d}} = (T_d - t_{{\rm el}_{\ell}})$ and $\dot{t}_{go_{d}}=-1$, we may write the error dynamics in time-to-go for the leader as
\begin{align*}
e_{l}(t) & = t_{go_{l}}-t_{go_{d}} = t_{go_{\ell}} - (T_d - {t_{{\rm el}_\ell}}),\\
\dot{e}_{\ell}(t) &= \dot{t}_{go_{\ell}} - \dot{t}_{go_{d}} = F_{\ell} + B_{\ell} a_{M_{\ell}} + 1.
\end{align*}
We use feedback linearization on the impact time error dynamics and choose the lateral acceleration of the leader interceptor in terms of the auxiliary control input $u_{\ell}$ as
\begin{align}
a_{M_{\ell}} = \frac{1}{B_{\ell}}\left(u_{\ell} - F_{\ell} - 1\right),
\end{align}
which reduces the time-to-go error dynamics of the leader to that of an integrator given by 
$\dot{e}_{\ell}(t) = u_{\ell}$.

To achieve target interception at the desired time $t=T_d$, we propose the auxiliary control input $u_l$ for the leader interceptor as
\begin{align}
\label{eq:tgo_single_integrator_leader_control_law}
u_{\ell} = \begin{cases}
-\dfrac{\eta_{\ell}}{t_{f_{\ell}} - t}(1-e^{-e_{\ell}(t)}), & \text{ if } t_0 \le t < t_{f_{\ell}} \\
0, & \text{ if } t \ge t_{f_{\ell}}
\end{cases}
\end{align}
with $\eta_{\ell} > 1$. Following \cite{Pal2020a}, it can be guaranteed that the control law in \eqref{eq:tgo_single_integrator_leader_control_law} leads to $e_{\ell}(t)$ converging to 0 at some time $t$ such that $t \le t_{f_{\ell}}$, where $t_{f_{\ell}}$ is the free-will arbitrary pre-specified time. This means that $t_{go_{\ell}}$ converges to $t_{go_{d}} = (T_d - t_{{\rm el}_{\ell}})$ within a finite time, $t_{f_{\ell}}$ using the following lateral acceleration control law
\begin{align}
a_{M_{\ell}} =  \begin{cases}
\frac{1}{B_{\ell}}\left[-\dfrac{\eta_{\ell}}{t_{f_{\ell}} - t}\left(  1-e^{-(t_{go_{l}}-t_{go_{d}})} \right) - F_{\ell} - 1\right], \\ \qquad \qquad\qquad\qquad\qquad     \qquad \text{ if } t_0 \le t < t_{f_{\ell}}  \\
\frac{1}{B_{\ell}}\left(- F_{\ell} - 1\right),  \qquad\qquad\qquad \text{ if } t \ge t_{f_{\ell}}
\end{cases}
\end{align}
and maintains this condition till target interception occurs.  

Next, we design the consensus law for the follower agents such that the control command for achieving consensus begins only after the leader has achieved convergence to its desired value, i.e. for $t\ge t_{f_{\ell}}$. Now, $\dot{t}_{go_{\ell}} = -1$ for all $t \ge t_{f_{\ell}}$. The dynamics of the time-to-go error $e_{i}(t) =t_{go_{i}} - t_{go_{\ell}}$ for the $i^{\rm th}$ follower is given by
\begin{align*}
\dot{e}_{i}(t) = \dot{t}_{go_{i}} - \dot{t}_{go_{\ell}} =  F_i + B_i a_{M_i} + 1,
\end{align*}
for $i\in\{1,2,\ldots,n\}$. The lateral acceleration command is written in terms of the auxiliary control command $u_i$ as follows
\begin{align}
\label{eq:single_integrator_control_law_am}
a_{M_{i}} = \frac{1}{B_i}\left(u_i - F_i - 1\right),
\end{align}
which reduces the error dynamics of each interceptor to that of a single integrator of the form $\dot{e}_{i}(t) = u_i$.

Writing $\bm{\delta} = [e_1(t),~e_2(t),\ldots,e_n(t)]^{T}$ and $\bm{u_f} = [u_1(t),~u_2(t),\ldots,u_n(t)]^{T}$, we design the consensus control law for the $n$ follower interceptors to achieve consensus to their leader's time-to-go within a free-will arbitrary pre-specified time $t_f$ as
\begin{align}
\label{eq:tgo_single_integrator_follower_consensus_law}
\bm{u_f} = 
\begin{cases}
\bm{0}, & \text{if}~ t_{0}\leq t < t_{f_{\ell}} \\
-\frac{\eta_f}{t_{f}-t}(\mathbf{1}_{n}-e^{-\bm{\mathcal{H}}\bm{\delta}}), & \text{if}~ t_{f_{\ell}}\leq t < t_{f} \\
\bm{0}, & \text{otherwise},
\end{cases}
\end{align}
where the gain $\eta_f > \frac{1}{\lambda_{\min}(\bm{\mathcal{H}})}$, and $\lambda_{\min}(\bm{\mathcal{H}})$ is the smallest eigenvalue of $\bm{\mathcal{H}}$. Note that this must be real as $\mathcal{H}$ is a symmetric matrix.

We next consider the Lyapunov candidate $V = \bm{\delta}^{T}\bm{\mathcal{H}}\bm{\delta}$. Its time  derivative may be obtained as 
\begin{align}
\dot{V} & = \dot{\bm{\delta}}^{T}\bm{\mathcal{H}}\bm{\delta}+\bm{\delta}^{T}\bm{\mathcal{H}}\dot{\bm{\delta}}  = \bm{\delta}^{T}\left( \bm{\mathcal{H}} + \bm{\mathcal{H}}^{T} \right)\dot{\bm{\delta}}\\  &= 2 \bm{\delta}^{T}\bm{\mathcal{H}}^{T}\dot{\bm{\delta}} = -\frac{2\eta_f}{t_{f}-t}\bm{\delta}^{T}\bm{\mathcal{H}}^{T}\left( \mathbf{1}_{n}-e^{-\bm{\mathcal{H}}\bm{\delta}} \right) \\
& =  -\frac{2\eta_f}{t_{f}-t}\left( \bm{\mathcal{H}}\bm{\delta}  \right)^{T}\left( \mathbf{1}_{n}-e^{-\bm{\mathcal{H}}\bm{\delta}} \right) \\
\label{ineq}
& \leq -\frac{2\eta_f}{t_{f}-t}\|\bm{\mathcal{H}}\bm{\delta}\|\left( 1-e^{-\|\bm{\mathcal{H}}\bm{\delta}\|} \right).
\end{align}
The above inequality is due to Lemma \ref{lemma:multi_var_inequality}. Further, we have used $\bm{\mathcal{H}} = \bm{\mathcal{H}}^{T}$ in the above derivation. If $\lambda_{\min}$ is the smallest eigenvalue of $\bm{\mathcal{H}}$, we have the following inequality
\begin{align}
\lambda_{\min}(\bm{\mathcal{H}})\|\bm{\delta}\|^{2} \leq \bm{\delta}^{T}\bm{\mathcal{H}}\bm{\delta} \leq \|\bm{\delta}\|\|\bm{\mathcal{H}}\bm{\delta}\|,
\end{align}
which results in
\begin{align}
\label{ineq1}
\lambda_{\min}(\bm{\mathcal{H}})\|\bm{\delta}\| \leq \|\bm{\mathcal{H}}\bm{\delta}\|\implies \|\bm{\delta}\|\leq \frac{\|\bm{\mathcal{H}}\bm{\delta}\|}{\lambda_{\min}(\bm{\mathcal{H}})}, 
\end{align}
since $\lambda_{\min}(\bm{\mathcal{H}})>0$. Using \eqref{ineq1}, the Lyapunov candidate may then be bounded by
\begin{align}
\begin{split}
V & = \bm{\delta}^{T}\bm{\mathcal{H}}\bm{\delta} 
  \leq \|\bm{\delta}\|\|\bm{\mathcal{H}}\bm{\delta}\| 
  \leq \frac{\|\bm{\mathcal{H}}\bm{\delta}\|^{2}}{\lambda_{\min}(\bm{\mathcal{H}})}.
\end{split}
\end{align}
From the above inequality, it follows that
$-\|\bm{\mathcal{H}}\bm{\delta}\|  \leq - \sqrt{ \lambda_{\min}(\bm{\mathcal{H}})V }$.
Based on Lemma \ref{lemma:single_var_inequality}, we may conclude that
\begin{align}
-\|\bm{\mathcal{H}}\bm{\delta}\|\left( {1}-e^{-\|\bm{\mathcal{H}}\bm{\delta}\|} \right) \leq -\sqrt{ \lambda_{\min}(\bm{\mathcal{H}}) V } \left( 1-e^{-\sqrt{ \lambda_{\min}(\bm{\mathcal{H}}) V }} \right).
\end{align}
This, combined with \eqref{ineq}, leads to the following inequality
\begin{align}
\dot{V} & \leq -\frac{2\eta_f \sqrt{ \lambda_{\min}(\bm{\mathcal{H}})V }}{t_{f}-t} \left( 1-e^{-\sqrt{ \lambda_{\min}(\bm{\mathcal{H}})V }} \right).
\end{align}
We define a new variable, $\xi$, for notational simplicity as $\xi = \sqrt{ \lambda_{\min}(\bm{\mathcal{H}}) V }$. The time derivative of $\xi$ may be written as
\begin{align*} 
\dot{\xi} & = \frac{\sqrt{ \lambda_{\min}(\bm{\mathcal{H}})} \dot{V} }{2\sqrt{ V }} \\
&\leq - \frac{\lambda_{\min}(\bm{\mathcal{H}})\eta_f}{t_{f}-t}\left( 1-e^{-\xi} \right)  = - \frac{\eta'}{t_{f}-t}\left( 1-e^{-\xi} \right), 
\end{align*}
where $\eta'=\lambda_{\min}(\bm{\mathcal{H}})\eta_f$. By Lemma \ref{lemma:fwat_convergence}, $\xi$ goes to zero at some $t \leq t_{f}$ if $\eta' > 1$, which implies $V$ also goes to zero at the same instant $t \leq t_{f}$ following the definition of $\xi$. Since $V$ is a positive definite function of $\bm{\delta}$, $V=0$ implies $\bm{\delta}=\bm{0}$ at that instant $t \leq t_{f}$. 

Therefore, the errors in time-to-go of the followers converge to $\bm{0}$ at a time $t$ in the interval $ t_{f_{\ell}} < t \le t_f $ when driven by the consensus protocol given by \eqref{eq:tgo_single_integrator_follower_consensus_law}. In other words, the time-to-go of the followers converges to $t_{go_{\ell}}$ at some time $t \le t_f$ where $t_f$ is the free-will arbitrary pre-specified time.

\begin{rem}
\label{rem:singularity_in_command}

Following \cite{Kumar2020}, it can be shown that there is no possibility of singularities occuring due to the leading angle $\theta_{M_i} \to 0$ as long as the consensus occurs before the interceptors converge on their respective collision courses. Hence, by suitable choice of $t_{f_l}$ and $t_f$ such that they are sufficiently less than the desired interception time $T_d$ as well as the gains $\eta_l$ and $\eta_f$, it can be ensured that $u_l$ and $\bm{u_f}$ become zero before the interceptors converge on their respective collision courses.
\end{rem}

\subsection{Salvo guidance with first-order autopilot}

Next, we propose a salvo guidance strategy for a group of interceptors with first-order autopilot dynamics, modeled by \eqref{eq:first_order_autopilot_dynamics}, which share information among each other over a leader-follower network. To incorporate the autopilot dynamics, we differentiate \eqref{eq:tgodot_estimate} with respect to time and use \eqref{eq:first_order_autopilot_dynamics} to obtain the following second-order time-to-go dynamics
\begin{align}
\label{eq:tgo_ddot_second_order} 
\ddot{t}_{go_i} &= \dot{F}_i + \dot{B}_ia_{M_i} + B_i\dot{a}_{M_i} \\
&= \dot{F}_i + \dot{B}_ia_{M_i} - \frac{B_ia_{M_i}}{\tau_i} + \frac{B_i}{\tau_i}a_{Mc_i}. 
\end{align}
The control command in this case is the autopilot command $a_{Mc_i}$, and the lateral acceleration of the interceptors evolves with time as per \eqref{eq:first_order_autopilot_dynamics}.

Using the following expressions and differentiating $F_i$ and $B_i$, we can derive the terms $\dot{F}_i$ and $\dot{B}_i$
\begin{align} 
\dot{\theta}_{M_i} &= \left(\dot{\gamma}_{M_i} - \dot{\theta}_i\right) = \left(\frac{a_{M_i}}{V_{M_i}} - \frac{V_{\theta_i}}{r_i}\right), \\
\dot{V}_{r_i} &= V_{M_i}\sin\theta_{M_i}\dot{\theta}_{M_i} = - \frac{a_{M_i}V_{\theta_i}}{V_{M_i}} + \frac{V_{\theta_i}^2}{r_i}, \\
\dot{V}_{\theta_i} &= -V_{M_i}\cos\theta_{M_i}\dot{\theta}_{M_i} = \frac{a_{M_i}V_{r_i}}{V_{M_i}} - \frac{V_{r_i}V_{\theta_i}}{r_i}.
\end{align}
After some algebraic manipulations, we get
\begin{align*}
\dot{F}_i&= \left( - \frac{a_{M_i}V_{\theta_i}}{V_{M_i}r_i} + \frac{V_{\theta_i}^2 - V_{r_i}^2 }{r_i^2} \right)t_{go_i} + \left(\frac{V_{r_i}}{r_i}\right)\dot{t}_{go_i}  \\
&~~- \left( \frac{V_{r_i}\theta_{M_i} + V_{\theta_i}}{V_{M_i}^2 K} \right)a_{M_i} + \frac{V_{r_i}V_{\theta_i}\theta_{M_i} + V_{\theta_i}^2}{r_i V_{M_i} K},\\
\dot{B}_i 
&= \frac{1}{V_{M_i}^3 K} \left(\theta_{M_i}V_{r_i}V_{M_i} + r_i a_{M_i} - V_{M_i}V_{\theta_i} \right).
\end{align*}
The proposed control law for the leader is discussed first. The dynamics of the time-to-go error $e_{l}(t) =t_{go_{l}} - t_{go_{d}} $ for the leader may be obtained as
\begin{align*} 
\dot{e}_{l}(t) &= \dot{t}_{go_{l}} - \dot{t}_{go_{d}}  
\implies \ddot{e}_{l}(t) = \ddot{t}_{go_{l}} - \ddot{t}_{go_{d}},
\end{align*}
where the desired time-to-go is $t_{go_{d}} = T_{d}-t_{{go}_{\ell}}$. Consequently, we have $\dot{t}_{go_{d}} = -1$ and $\ddot{t}_{go_{d}} = 0$. Next, we design the lateral control command, $a_{Mc_{\ell}}$, in terms of an auxiliary control input, $u_{\ell}$, as
\begin{align}
a_{Mc_{\ell}} = \frac{\tau_{\ell}}{B_{\ell}}\left[ u_{\ell} - \dot{F}_{\ell} - \left( \dot{B}_{\ell} - \frac{B_{\ell}}{\tau_{\ell}} \right) a_{M_{\ell}}  \right]. 
\end{align}
With the above auxiliary control input, the error dynamics corresponding to leader interceptor results in the following double-integrator form
\begin{align}
\label{eq:second_order_tgo_dynamics_leader}
\ddot{e}_{\ell}(t) = u_{\ell}.
\end{align}
It can further be expressed in the state-space form, with $\delta_{\ell} = e_{\ell}$ and $\nu_{\ell} = \dot{e}_{\ell}$, as $\dot{\delta}_{\ell} = \nu_{\ell},~\dot{\nu}_{\ell} = u_{\ell}$. For achieving target interception at the desired time $T_d$, we propose the following control law for the leader interceptor
\begin{align}
\label{eq:second_order_control_leader}
u_{\ell} = \begin{cases}
-\delta_{\ell} - \dfrac{\partial \psi_1}{\partial t} - \nu_{\ell}\dfrac{\partial \psi_1}{\partial \delta_{\ell}} - \dfrac{\eta_{\ell_2}\left( 1-e^{-z_{\ell}} \right)}{t_{f_{\ell}} - t},~t\in [t_0,\,t_{f_{\ell}]} \\
0, ~\text{otherwise,}
\end{cases}
\end{align}
where $\psi_1(\delta_{\ell},t)$ and $z_{\ell}$ are defined as $\psi_1(\delta_{\ell},t) = -\dfrac{\eta_{\ell}}{t_{f_{\ell}} - t}\left( 1-e^{-\delta_{\ell}} \right),~ z_{\ell} = \nu_{\ell} + \psi_1$, respectively, with the parameters $\eta_{\ell},~ \eta_{\ell_2} \ge 1$. The free-will arbitrary time within which the leader's time-to-go converges to its desired trajectory is denoted by $t_{f_{\ell}}$. Following \cite{Pal2020a} (Section 4.2), the control law given in \eqref{eq:second_order_control_leader} ensures that $\delta_{\ell} \to 0$ and $\nu_{\ell} \to 0$ at some $t < t_{f_{\ell}}$. Hence, it is established that the leader time-to-go converges to the desired $t_{go_{d}} = T_{d}-t_{{\rm el}_{\ell}}$, and its derivative converges to $-1$ within the free-will arbitrarily specified time, $t_{f_{\ell}}$ using the following control law for the autopilot command $a_{Mc_{\ell}}$ as
\begin{align*}
a_{Mc_{\ell}} =  \frac{\tau_{\ell}}{B_{\ell}}\left[ u_{\ell} - \dot{F}_{\ell} - \left( \dot{B}_{\ell} - \frac{B_{\ell}}{\tau_{\ell}} \right) a_{M_{\ell}}  \right],
\end{align*}
where the terms $u_{\ell}$ and $\psi_1(\delta_{\ell},t)$ are defined as
\begin{align*}
&u_{\ell} =  \begin{cases}
-\delta_{\ell} - \dfrac{\partial \psi_1}{\partial t} - \nu_{\ell}\dfrac{\partial \psi_1}{\partial \delta_{\ell}} - \dfrac{\eta_{\ell_2}\left( 1-e^{-z_{\ell}} \right)}{t_{f_{\ell}} - t},~ t\in[t_0,\,t_{f_{\ell}]} \\
0, ~\text{otherwise,}
\end{cases} \\
&\psi_1(\delta_{\ell},t) =  -\dfrac{\eta_{\ell}}{t_{f_{\ell}} - t}\left( 1-e^{-\delta_{\ell}} \right),~ z_{\ell} = \nu_{\ell} + \psi_1
\end{align*}
with $\eta_{\ell},~ \eta_{\ell_2} \ge 1$.

The salvo guidance for the follower interceptors is designed considering an undirected communication topology among themselves and unidirectional communication from the leader communicating to one or more followers. As before, the consensus law for the followers is developed such that the control command initiates only after convergence of the leader's time-to-go to the desired $t_{go} = T_d - t_{{\rm el}_i}$. Thus, $\dot{t}_{go_{\ell}} = -1$ and $\ddot{t}_{go_{\ell}} = 0$ and the dynamics of time-to-go error $e_{i}(t)  =t_{go_{i}} - t_{go_{\ell}}$ for the $n$ followers is obtained as
\begin{align*}  
\dot{e}_{i}(t) &= \dot{t}_{go_{i}} - \dot{t}_{go_{\ell}} =  \dot{t}_{go_{i}} + 1  \\
  &= \ddot{t}_{go_{i}} - \ddot{t}_{go_{\ell}} = \ddot{t}_{go_{i}},
\end{align*}
for $i\in\{1,2,\ldots,n\}$.
Using \eqref{eq:tgo_ddot_second_order}, the error dynamics of the $i^{\rm th}$ follower results in the following second-order form
\begin{align}
\ddot{e}_{i}(t) = \dot{F}_{i} + \left( \dot{B}_{i} - \frac{B_{i}}{\tau_{i}} \right)a_{M_{i}} + \frac{B_{i}}{\tau_{i}}a_{{Mc_{i}}}.
\end{align}
The lateral control command, $a_{Mc_i}$, is designed in terms of an auxiliary control input, $u_i$, as
\begin{align}
a_{Mc_{i}} = \frac{\tau_{i}}{B_{i}}\left[ u_{i} - \dot{F}_{i} - \left( \dot{B}_{i} - \frac{B_{i}}{\tau_{i}} \right) a_{M_{i}}  \right].  
\end{align}
Accordingly, the error dynamics results in the double-integrator dynamics form
\begin{align}
\label{eq:second_order_tgo_dynamics_follower}
\ddot{e}_{i}(t) = u_{i},
\end{align}
which may be rewritten in the state-space form, in terms of $\bm{\delta} = [e_{1}, e_{2}, \dots, e_{n}] \in \mathbb{R}^{n}$, $\bm{\nu} = [\dot{e}_{1},\dot{e}_{2},\dots,\dot{e}_{n}] \in \mathbb{R}^{n}$, and $\bm{u_f}=[u_{1},u_{2},\dots,u_{n}]\in \mathbb{R}^{n}$ as
\begin{align}
\label{eq:second_order_follower_dynamics}
\dot{\bm{\delta}} &= \bm{\nu},\qquad
\dot{\bm{\nu}} = \bm{u_f}.
\end{align}
To design the consensus law for the follower interceptors using back-stepping, we apply the following variable change $\bm{z} = \bm{\nu} + \bm{\phi}$, where we define $\bm{\phi}(t,\bm{\delta}) = \dfrac{\eta_{f}}{t_{f}-t}\left( \mathbf{1}_{n} - e^{-\bm{\mathcal{H}}\bm{\delta}} \right)$ with $\eta_f > \dfrac{1}{\lambda_{\min}(\bm{\mathcal{H}})}$. The time derivative of $\bm{z}$ is obtained as
\begin{align*}
\dot{\bm{z}} & = \dot{\bm{\nu}} + \frac{ \partial \bm{\phi} }{ \partial \bm{\delta} }\bm{\nu} + \frac{ \partial \bm{\phi} }{ \partial t } = \bm{u_f} + \frac{ \partial \bm{\phi} }{ \partial \bm{\delta} }\bm{\nu} + \frac{ \partial \bm{\phi} }{ \partial t },
\end{align*}
where the terms $\dfrac{ \partial \bm{\phi} }{ \partial \bm{\delta} }$ and $\dfrac{ \partial \bm{\phi} }{ \partial t }$ are defined as
\begin{align*}
\frac{ \partial \bm{\phi} }{ \partial \bm{\delta} }  & = \frac{\eta_{f}}{t_{f} - t} \text{diag}\left( e^{-\bm{\mathcal{H}}\bm{\delta}} \right)\bm{\mathcal{H}},~~
\frac{ \partial \bm{\phi} }{ \partial t }  = \frac{\eta_{f}}{(t_{f}-t)^{2}}\left( \mathbf{1}_{n} - e^{-\bm{\mathcal{H}}\bm{\delta}} \right).
\end{align*}
We propose the following consensus law for the follower interceptors
\begin{align}
\label{eq:second_order_follower_consensus_law}
\bm{u_f} = \begin{cases}
\bm{0}, & \text{if}~ t_{0}\leq t < t_{f_{\ell}}, \\
-\frac{ \partial \bm{\phi} }{ \partial \bm{\delta} }\bm{\nu} - \frac{ \partial \bm{\phi} }{ \partial t } - \frac{\eta_{f_{2}}}{t_{1} - t} (\mathbf{1}_{n} - e^{-\bm{z}}), & \text{if}~ t_{f_{\ell}}\leq t < t_{1} \\
-\frac{ \partial \bm{\phi} }{ \partial \bm{\delta} }\bm{\nu} - \frac{ \partial \bm{\phi} }{ \partial t }, & \text{if}~ t_{1} \leq t < t_{f} \\
\bm{0}, & \text{otherwise},
\end{cases}
\end{align}
with $\eta_{f_{2}} > 1$. Hence, in the interval $t_{f_{\ell}} \leq t < t_{1}$, \eqref{eq:second_order_follower_consensus_law} results in the following closed-loop dynamics
\begin{align}
\dot{\bm{z}} = - \frac{\eta_{f_{2}}}{t_{1} - t} (\mathbf{1}_{n} - e^{-\bm{z}}).
\end{align}

We consider the  Lyapunov candidate $V_{z} = \bm{z}^{T}\bm{z}$, whose time derivative may be obtained as
\begin{align} 
\dot{V}_{z} & = 2 \bm{z}^{T}\dot{\bm{z}}
   = -2 \frac{\eta_{f_{2}}}{t_{1}-t}\bm{z}^{T}(\mathbf{1}_{n}-e^{-\bm{z}})\\
   &\leq -2 \frac{\eta_{f_{2}}}{t_{1}-t}\|\bm{z}\|(1-e^{-\|\bm{z}\|})\\  
   &\leq -2 \frac{\eta_{f_{2}}}{t_{1}-t}\sqrt{ V_{z} }(1 - e^{-\sqrt{ V_{z} }}), 
\end{align}
where the last two inequalities follow from Lemmas \ref{lemma:single_var_inequality} and \ref{lemma:multi_var_inequality} along with the fact that $V_{z} = \|\bm{z}\|^{2}$. Consider $\xi = \sqrt{ V_{z} }$ and observe that its time derivative is given by
\begin{align}
\begin{split}
\dot{\xi} & = \frac{\dot{V}_{z}}{2\sqrt{ V_{z} }} \leq -\frac{\eta_{f_{2}}}{t_{1}-t}(1 - e^{-\xi}).
\end{split}
\end{align}
By Lemma \ref{lemma:fwat_convergence}, if $\eta_{f_{2}} > 1$, $\xi$ goes to zero at some time $t$ in the interval $t_{f_{\ell}} \le t < t_1$, which implies that $V_z = \xi^2$ also goes to zero at the same instant. Since $V_z$ is a positive definite function of $\bm{z}$, $\bm{z} \to \mathbf{0}$ as $t \to t_{1}$. Hence, $\bm{\nu}(t)=-\bm{\phi}$ for all $t \geq t_{1}$. Thereafter, the system reduces to the following single integrator dynamics
\begin{align}
\dot{\bm{\delta}} = -\frac{\eta_{f}}{t_{f}-t}\left( \mathbf{1}_{n} - e^{-\bm{\mathcal{H}}\bm{\delta}} \right), & ~\forall~t \geq t_{1}.
\end{align}
Finally, using the results for the system with single integrator dynamics obtained in the case of ideal autopilot earlier, it follows that $\bm{\delta} \to \bm{0}$ as $t \to t_{f}$ if $\eta_f > \frac{1}{\lambda_{\min}(\bm{\mathcal{H}})}$. Since $\bm{\nu} = -\bm{\phi}$ for all $t \geq t_{1}$, as $\bm{\delta}$ converges to $\bm{0}$, it implies that $\bm{\phi}$ converges to $\bm{0}$, and hence $\bm{\nu} \to \bm{0}$ as $t \to t_{f}$. Equivalently, it follows that $\bm{\delta} = \bm{0}$ and $\bm{\nu}=\bm{0}$ for all $t \geq t_{f}$.

Therefore, we have shown that for a system of interceptors in a leader-follower communication framework, with double integrator time-to-go dynamics given by \eqref{eq:second_order_follower_dynamics}, the follower time-to-go error, $\bm{\delta}$, and its derivative, $\bm{\nu}$, converge to $\bm{0}\in \mathbb{R}^n$ within a fixed time which is bounded in the interval  $t_{f_{\ell}} \le t < t_f - t_{f_{\ell}}$. 
The settling times $t_{f_{\ell}}$ and $t_f$ must be chosen such that  $t_{f_{\ell}} < t_f$ with adequate margin between the two times to allow sufficient maneuvering time for the followers. 

\section{Simulation results and discussions}
\label{sec:simulation}
\begin{figure}[tb]
	\centering
	\includegraphics[width=0.4\textwidth]{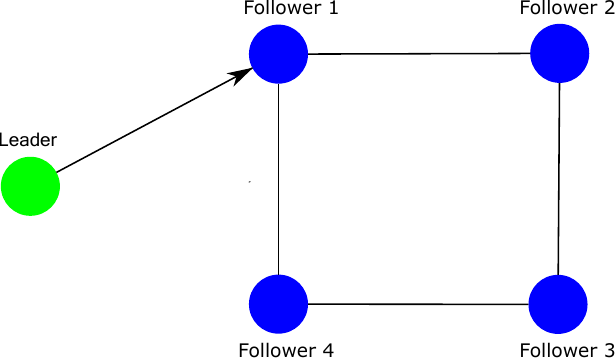}
	\caption{Communication topology with 1 leader and 4 followers for simulations.}
	\label{fig:network_graph}
\end{figure}

Simulations to evaluate the efficacy of the proposed salvo strategies with the system lag approximated as ideal autopilot and first-order autopilot are presented in this section. We consider the same communication network in both scenarios, which is illustrated in Fig.~\ref{fig:network_graph}. There are 4 follower interceptors which share information with each other over an undirected cycle graph. The leader interceptor communicates with one of the followers only over a directed edge unidirectionally. 

The followers' graph Laplacian, $\bm{\mathcal{L}_{f}}$, and the leader incidence matrix $\bm{\mathcal{B}}$, for the network in Fig.~\ref{fig:network_graph}, are
\begin{align*}
\bm{\mathcal{L}_{f}} = \begin{bmatrix}
2 & -1 & 0 & -1 \\
-1 & 2 & -1 & 0 \\
0 & -1 & 2 & -1 \\
-1 & 0 & -1 & 2
\end{bmatrix}; & ~~~  
\bm{\mathcal{B}} = \begin{bmatrix}
1 & 0 & 0 & 0 \\
0 & 0 & 0 & 0 \\
0 & 0 & 0 & 0 \\
0 & 0 & 0 & 0
\end{bmatrix}.
\end{align*}
The interaction matrix $\bm{\mathcal{H}}=\bm{\mathcal{L}_{f}}+\bm{\mathcal{B}}$ has $\lambda_{\min}(\bm{\mathcal{H}}) = 0.186$. We present different initial conditions for the simulations in order to assess the efficiency and robustness of the proposed salvo guidance strategy, which are summarized in Table~\ref{tab:tgo_simulation_initial_conditions}. Set 1 is used to assess the effect of different heading angle errors $\theta_{M} = \gamma_{M} - \theta$ using different values of initial azimuth $\theta$ and flight path angle $\gamma_{M}$. Set 2 demonstrates the effects of different starting positions of the interceptors at different initial ranges, while Set 3 is used to assess the effects of different speeds of the interceptors. The origin of the coordinate system is considered to be at the stationary target for all the simulations.

\begin{table*}[t]
\centering
\caption{Table of initial conditions used for simulations.}
\label{tab:tgo_simulation_initial_conditions} 
\begin{tabular}{l l c c c c c}
\hline
\multirow{2}{*}{Data} & \multirow{2}{*}{Parameters} & \multicolumn{5}{c}{Interceptor} \\ \cline{3-7} 
 &  & Leader & Follower 1 & Follower 2 & Follower 3 & Follower 4 \\ \hhline{ - - - - - - -}
\multirow{4}{*}{Set 1} 
 & $r_0$ (\si{m}) & 5000 & 5000 & 5000 & 5000 & 5000 \\
 & $V_{M}$ (\si{m/s}) & 200 & 200 & 200 & 200 & 200 \\
 & $\theta_0$ & 45\textdegree & 0\textdegree & -45\textdegree & 60\textdegree & 120\textdegree \\
 & $\gamma_0$ & 90\textdegree & 60\textdegree & -30\textdegree & 30\textdegree & 70\textdegree \\ \hline
 \multirow{4}{*}{Set 2} 
 & $r_0$ (\si{m}) & 10500 & 11000 & 9800 & 10000 & 9500 \\
 & $V_{M}$ (\si{m/s}) & 200 & 200 & 200 & 200 & 200 \\
 & $\theta_0$ & -30\textdegree & -60\textdegree & -45\textdegree & -80\textdegree & -90\textdegree \\
 & $\gamma_0$ & -45\textdegree & -90\textdegree & -30\textdegree & -60\textdegree & -45\textdegree \\ \hline
  \multirow{4}{*}{Set 3} 
 & $r_0$ (\si{m}) & 6000 & 6000 & 600 & 6000 & 6000 \\
 & $V_{M}$ (\si{m/s}) & 630 & 630 & 600 & 570 & 594 \\
 & $\theta_0$ & -40\textdegree & -50\textdegree & -60\textdegree & -30\textdegree & -70\textdegree \\
 & $\gamma_0$ & 0\textdegree & -5\textdegree & 0\textdegree & 25\textdegree & 10\textdegree \\ \hline
\end{tabular}
\end{table*}

\subsection{Results for salvo guidance with ideal autopilot}
\label{subsec:results_ideal_autopilot}

In this section, we discuss the simulations results for salvo guidance with ideal autopilot using the initial conditions given in Table~\ref{tab:tgo_simulation_initial_conditions}.  The lateral acceleration command, $a_{M_i}$, for each interceptor is bounded within \SI{300}{m/s^2}.  

\subsubsection{Effect of heading angle errors}

Each interceptor is initially located at the same radial distance of \SI{5000}{m} from the target but at different azimuth angles, $\bm{\theta}_0 = [\SI{45}{\degree},~\SI{0}{\degree},~\SI{-45}{\degree},~\SI{60}{\degree},~\SI{120}{\degree}]^T$. Their constant identical speed is \SI{200}{m/s} with different initial heading angles $\bm{\gamma}_0 = [\SI{90}{\degree},~\SI{60}{\degree},~\SI{-30}{\degree},~\SI{30}{\degree},~\SI{70}{\degree}]^T$. A desired time of interception is chosen as \SI{30}{s}. The free-will arbitrary time, $t_{f_{\ell}}$, within which the leader is required to converge to the desired value is \SI{5}{s}, while the free-will arbitrary time, $t_f$ required for consensus by the followers to the leader's $t_{go}$ is \SI{25}{s}. The parameters $\eta_{\ell}$ and $\eta_f$ are chosen as 2 and 10.73, respectively.

\begin{figure*}[tb]
    \centering
    \begin{subfigure}[t]{0.32\linewidth}
    	\includegraphics[width=\textwidth]{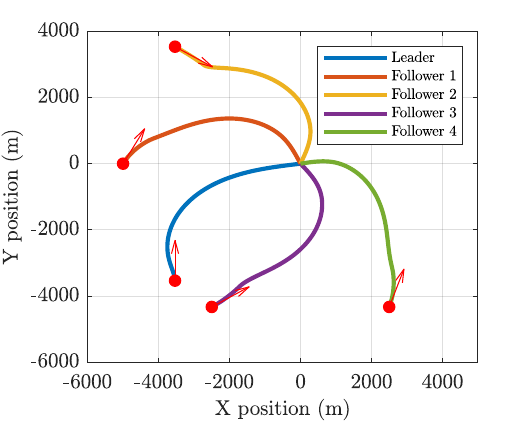}	
    	\caption{Trajectory.}
	    \label{fig:single_integrator_xy}
    \end{subfigure}
    \begin{subfigure}[t]{0.32\linewidth}
    	\includegraphics[width=\textwidth]{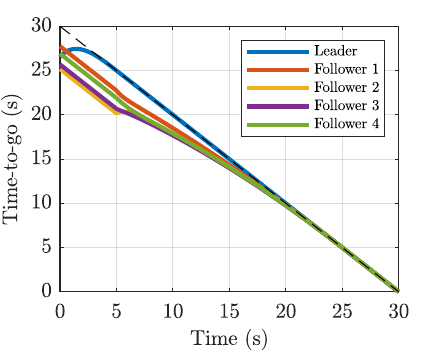}
    	\caption{Time-to-go.}
    	\label{fig:single_integrator_tgo}
    \end{subfigure}
    \begin{subfigure}[t]{0.32\linewidth}
    	\includegraphics[width=\textwidth]{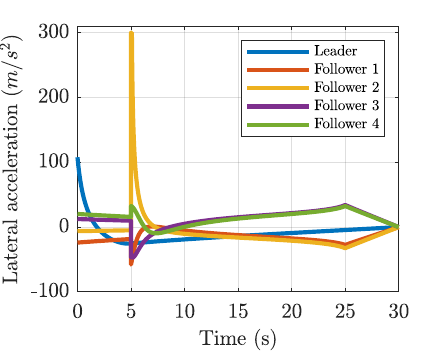}
    	\caption{Lateral acceleration command.}
    	\label{fig:single_integrator_am}
    \end{subfigure}
    \begin{subfigure}[t]{0.32\linewidth}
    \includegraphics[width=\textwidth]{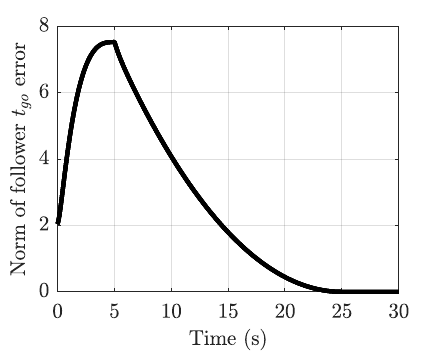}
	\caption{Follower consensus error.}
	\label{fig:single_integrator_error_f}
    \end{subfigure}  
    \begin{subfigure}[t]{0.32\linewidth}
    \includegraphics[width=\textwidth]{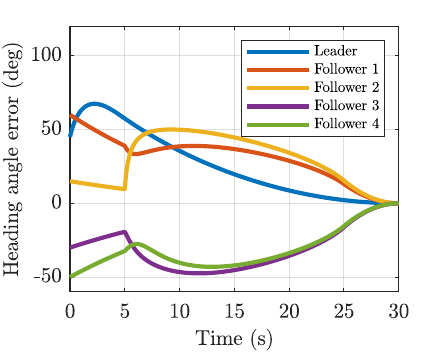}
	\caption{Heading angle error.}
	\label{fig:single_integrator_thetaM}
    \end{subfigure}
    \caption{Simulation results of cooperative guidance strategy for interceptors with ideal interceptor autopilot.}
    \label{fig:single_integrator}
\end{figure*}

Successful interception of the target by each interceptor is demonstrated in Fig.~\ref{fig:single_integrator_xy}, which shows the trajectories of the interceptors. The initial positions of the interceptors are marked by red dots, while the arrowheads at respective initial locations indicate directions of their initial velocity vectors. The evolution of $t_{go}$ in Fig.~\ref{fig:single_integrator_tgo} shows that  consensus is achieved by the follower interceptors to the leader's $t_{go}$ value well within \SI{10}{s}. As observed in the Fig.~\ref{fig:single_integrator_tgo}, the initial time-to-go for all the interceptors is less than \SI{30}{s} with the maximum being \SI{27.74}{s}. However, our guidance scheme ensures that the time-to-go for the leader converges to desired value within $t_{f_{\ell}} = \SI{5}{s}$, and since all the followers converge to the leader's time-to-go within \SI{25}{s} (as shown in Fig.~\ref{fig:single_integrator_error_f}), simultaneous interception of target occurs at the desired impact time of \SI{30}{s}. It may also be observed that the follower interceptors start their maneuver after $t=t_{f_{\ell}} = \SI{5}{s}$.  

The lateral acceleration command is depicted in Fig.~\ref{fig:single_integrator_am}, which shows that the lateral acceleration bound is reached for the follower 2 at the beginning of the trajectory, which is a consequence of the large time-to-go error for the follower 2 at the time of consensus control engagement. However, it quickly reduces to \SI{10}{m/s^2} within \SI{1.5}{s}. An important observation from Fig.~\ref{fig:single_integrator_am} is that the lateral acceleration command for each interceptor reduces to zero smoothly as they approach the target interception at \SI{30}{s}. This desirable feature is due to the convergence of $\dot{t}_{go}$ to $-1$ within a finite time before target interception. The headings of the interceptors also converge to zero, as shown in Fig.~\ref{fig:single_integrator_thetaM}, which indicates that interception of the target occurs with maximum impact velocity. This maximizes the damage to the target. 

\subsubsection{Effect of the choice of free-will arbitrary settling time $t_f$}

\begin{figure*}[tb]
    \centering
    \begin{subfigure}[t]{0.32\linewidth}
    	\includegraphics[width=\textwidth] {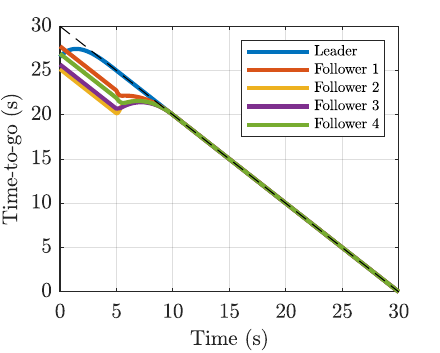}
    	\caption{Time-to-go.}
    	\label{fig:single_integrator_smaller_tf_tgo}
    \end{subfigure}
    \begin{subfigure}[t]{0.32\linewidth}
    	\includegraphics[width=\textwidth]{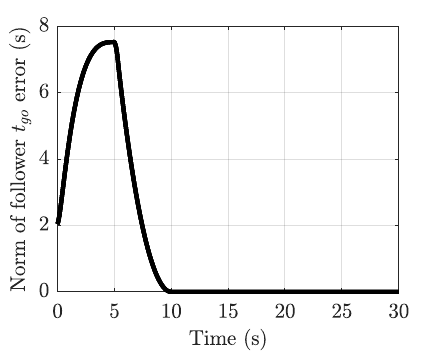}
    	\caption{Norm of follower consensus error.}
    	\label{fig:single_integrator_smaller_tf_error_f}
    \end{subfigure}
    \begin{subfigure}[t]{0.32\linewidth}
    	\includegraphics[width=\textwidth]{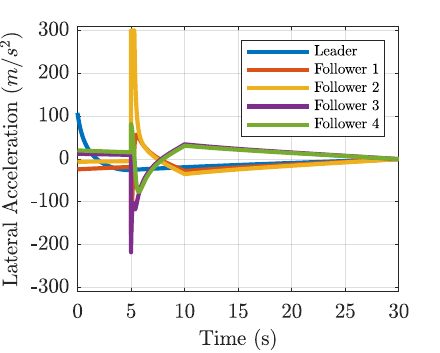}
    	\caption{Lateral acceleration command.}
    	\label{fig:single_integrator_smaller_tf_am}
    \end{subfigure}  
    \caption{Simulation results with reduced bound on settling time $t_f$.}
    \label{fig:single_integrator_smaller_tf}
\end{figure*}

Next, we choose the free-will arbitrary settling time, $t_f$, as \SI{10}{s} which is much smaller than the \SI{25}{s} selected in the previous case. In this case also, the desired impact time is achieved as shown in Fig.~\ref{fig:single_integrator_smaller_tf_tgo}. We observe in Fig.~\ref{fig:single_integrator_smaller_tf_error_f} that the consensus is achieved as expected within a smaller settling time $t_f = \SI{10}{s}$. However, Fig.~\ref{fig:single_integrator_smaller_tf_am} shows that the reduction in settling time causes an increase in the lateral acceleration commands on the follower interceptors at the beginning of the consensus control. The increase in saturation duration of the follower 2 acceleration command can be clearly observed when compared with Fig.~\ref{fig:single_integrator_am}. Further reduction in $t_f$ would lead to command saturation for the other interceptors as well. An increase in $t_f$ is possible up to $T_d = \SI{30}{s}$ since an increase beyond such value would not lead to a consensus before target interception.

\subsubsection{Effect of the choice of desired impact time $T_d$}

\begin{figure*}[tb]
    \centering
    \begin{subfigure}[t]{0.32\linewidth}
    	\includegraphics[width=\textwidth] {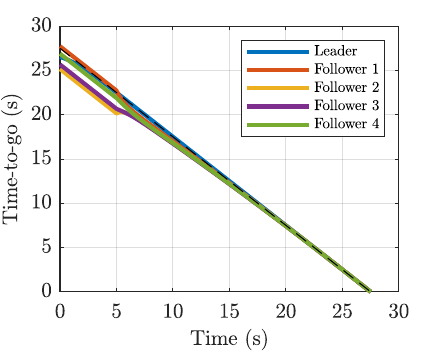}
    	\caption{Time-to-go.}
    	\label{fig:single_integrator_smaller_Td_tgo}
    \end{subfigure}
    \begin{subfigure}[t]{0.32\linewidth}
    	\includegraphics[width=\textwidth] {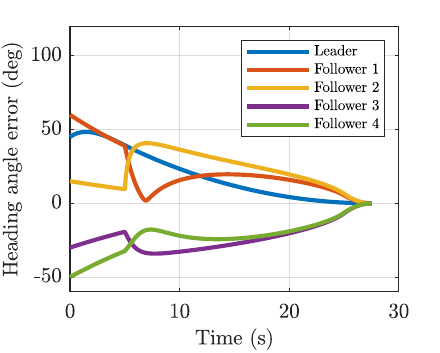}
    	\caption{Heading angle error.}
    	\label{fig:single_integrator_smaller_Td_thetaM}
    \end{subfigure}
    \begin{subfigure}[t]{0.32\linewidth}
    	\includegraphics[width=\textwidth] {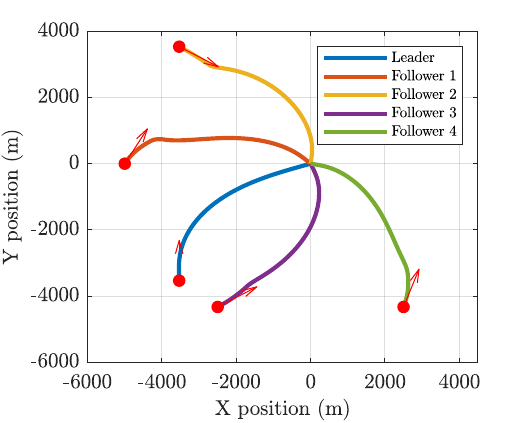}
    	\caption{Trajectory.}
    	\label{fig:single_integrator_smaller_Td_xy}
    \end{subfigure}  
    \caption{Simulation results with reduced desired impact time $T_d$.}
    \label{fig:single_integrator_smaller_Td}
\end{figure*}

We choose the desired impact time, $T_d$, as \SI{27.52}{s} which is smaller than \SI{30}{s} selected in the original case (with $t_f = \SI{25}{s}$). Figure~\ref{fig:single_integrator_smaller_Td_tgo} shows that the interceptors are able to intercept the target simultaneously at the desired impact time. However, it was seen that further reduction in impact time is not possible beyond this value. This is because, as observed in Fig.~\ref{fig:single_integrator_smaller_Td_thetaM}, the heading angle error $\theta_M$ of Follower 1 is close to 0, and a further reduction in $T_d$ would lead to $\theta_M=0$. 
This violates the condition given in Remark \ref{rem:singularity_in_command}. Comparing the trajectories of Follower 1 shown in Fig.~\ref{fig:single_integrator_smaller_Td_xy} with the original case shown in Fig.~\ref{fig:single_integrator_xy}. It may be clearly observed that the trajectory of follower 1 is rather flat and is close to a straight line. This means that for the particular choice of the position, velocity, and maximum achievable control input, the time-to-go for follower 1 has reached very close to its minimum value and cannot be reduced further. Hence, there is a lower bound on the desired impact time, $T_d$, depending on the operating conditions, such as the interceptors' velocities, initial positions, maximum achievable control inputs, and other physical constraints.

\subsubsection{Effect of different initial positions}

At the initial time, $t_0$, each interceptor is located at different radial distances $\vec{r_0} = [10500,~11000,~9800,~10000,~9500]$\si{m} from the target and at different azimuth angles, $\bm{\theta}_0 = [-30^{\circ},~-60^{\circ},~-45^{\circ}, ~-80^{\circ}, ~-90^{\circ}]^T$. Their constant identical speed is \SI{200}{m/s} with different initial heading angles $\bm{\gamma}_0 = [-45^{\circ},~-90^{\circ}, ~-30^{\circ},~-60^{\circ}, ~-45^{\circ}]^T$. The desired time of interception is chosen as $T_d = \SI{65}{s}$. The free-will arbitrary time, $t_{f_{\ell}}$, within which the leader is required to converge to the desired value is \SI{5}{s}, while the free-will arbitrary time, $t_f$, required for consensus by the followers to the leader's $t_{go}$ is \SI{50}{s}. The parameters $\eta_{\ell}$ and $\eta_f$ are chosen as 3 and 10.73, respectively. These parameters are similar to the previous case since they are mainly selected based on the particular communication topology at hand. 

\begin{figure*}[tb]
    \centering
    \begin{subfigure}[t]{0.32\linewidth}
    	\includegraphics[width=\textwidth] {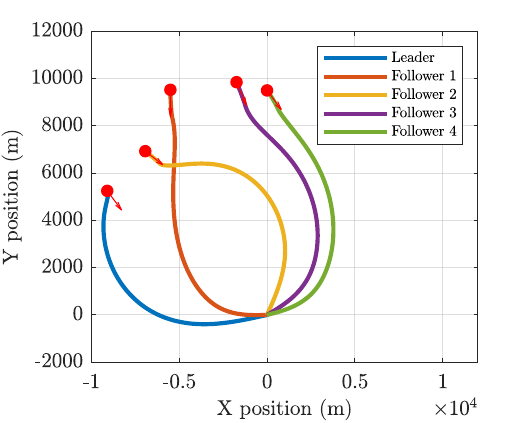}
    	\caption{Trajectory.}
    	\label{fig:tgo_single_interator_set2_xy}
    \end{subfigure}
    \begin{subfigure}[t]{0.32\linewidth}
    	\includegraphics[width=\textwidth] {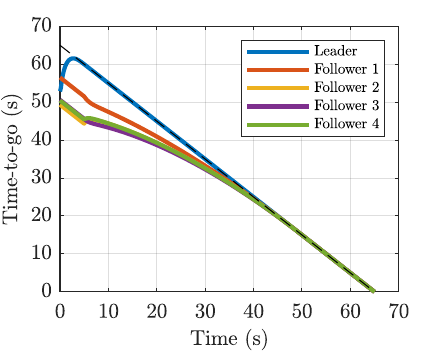}
    	\caption{Time-to-go.}
    	\label{fig:tgo_single_interator_set2_tgo}
    \end{subfigure}
    %
    %
    \begin{subfigure}[t]{0.32\linewidth}
    	\includegraphics[width=\textwidth] {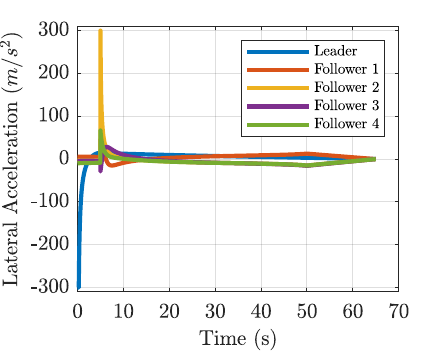}
    	\caption{Lateral acceleration command.}
    	\label{fig:tgo_single_interator_set2_am}
    \end{subfigure}  
    %
    \caption{Cooperative guidance strategy for interceptors  with ideal autopilot starting at different locations.}
    \label{fig:tgo_single_interator_set2}
\end{figure*}

Figure~\ref{fig:tgo_single_interator_set2_xy} shows that the proposed salvo strategy is able to achieve interception in this case too. Target interception occurs simultaneously by all the interceptors at the desired impact time $T_d = \SI{65}{s}$ as evident from Fig.~\ref{fig:tgo_single_interator_set2_tgo}. The norm of the follower consensus error converges to 0 within the selected $t_f = \SI{50}{s}$ which implies that the time-to-go of the followers converges to their leader's time-to-go within the desired $t_f = \SI{50}{s}$. It may be observed from Fig.~\ref{fig:tgo_single_interator_set3_am} that the lateral acceleration command for the leader and follower 2 saturate for a very brief duration during their respective initial phases, but subsequently reduce to smaller values soon enough. 

\subsubsection{Effect of different interceptor speeds}

At the initial instant, $t_0$, each interceptor is located at the same radial distance of $\SI{6000}{m}$ from the target but at different azimuth angles, $\bm{\theta}_0 = [-40{^\circ},~-50{^\circ},~-60{^\circ},~-30{^\circ}, ~-70{^\circ}]^T$. They have different speeds $\vec{v_{M}} = [630,~630,~600,~570,~594]^{T}$\si{m/s} with different initial heading angles $\bm{\gamma}_0 = [0{^\circ},~-5{^\circ},~0{^\circ},~25{^\circ},~10{^\circ}]^T$. The desired time of interception is chosen as $T_d = \SI{12}{s}$. The free-will arbitrary time, $t_{f_{\ell}}$, within which the leader is required to converge to the desired $t_{go_d} = 12 - t$ curve is \SI{2}{s}, while the free-will arbitrary time, $t_f$, required for consensus by the followers to the leader's $t_{go}$ is \SI{10}{s}. The parameters $\eta_{\ell}$ and $\eta_f$ are chosen as 3 and 10.73, respectively.

\begin{figure*}[tb]
    \centering
    \begin{subfigure}[t]{0.32\linewidth}
    	\includegraphics[width=\textwidth] {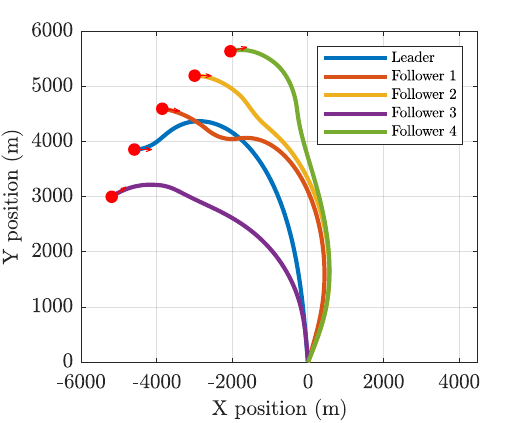}
    	\caption{Trajectory.}
    	\label{fig:tgo_single_interator_set3_xy}
    \end{subfigure}
    \begin{subfigure}[t]{0.32\linewidth}
    	\includegraphics[width=\textwidth] {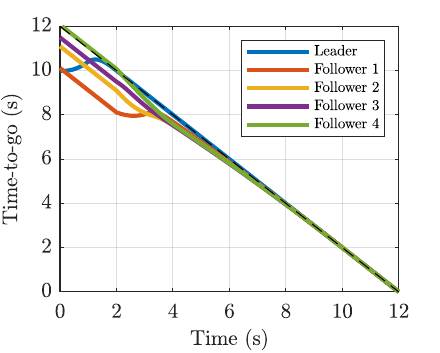}
    	\caption{Time-to-go.}
    	\label{fig:tgo_single_interator_set3_tgo}
    \end{subfigure}
    \begin{subfigure}[t]{0.32\linewidth}
    	\includegraphics[width=\textwidth] {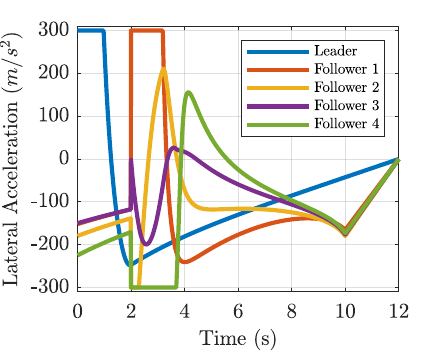}
    	\caption{Lateral acceleration command.}
    	\label{fig:tgo_single_interator_set3_am}
    \end{subfigure}
    %
    %
    \caption{Cooperative guidance strategy for interceptors with ideal autopilot and different velocities.}
    \label{fig:tgo_single_interator_set3}
\end{figure*}

Figure~\ref{fig:tgo_single_interator_set3_xy} shows that the proposed salvo strategy is able to achieve interception in this case with different interceptor velocities also. Target interception by all the interceptors occurs simultaneously at the desired impact time $T_d = \SI{12}{s}$, as evident from Fig.~\ref{fig:tgo_single_interator_set3_tgo}. As indicated in Fig.~\ref{fig:tgo_single_interator_set3_tgo}, the time-to-go of the followers converge to that of their leader within $t_f = \SI{10}{s}$. It may be observed from Fig.~\ref{fig:tgo_single_interator_set3_am} that the lateral acceleration commands for the leader and the followers 1, 2, and 4 saturate for some duration during their initial time instants, but subsequently reduce to smaller values very soon. 

\subsection{Results for salvo guidance with first-order autopilot}

In this subsection, we discuss the simulation results for salvo guidance with a first-order autopilot using the initial conditions given in Set 1 of  Table~\ref{tab:tgo_simulation_initial_conditions}. At the initial instant, $t_0$, each interceptor is located at the same radial distance of \SI{5000}{m} from the target but at different azimuth angles, $\bm{\theta}_0 = [\SI{45}{\degree},~\SI{0}{\degree}, ~\SI{-45}{\degree},~\SI{60}{\degree},~\SI{120}{\degree}]^T$. Their constant identical speed is \SI{200}{m/s} with different initial heading angles $\bm{\gamma}_0 = [\SI{90}{\degree}, ~\SI{60}{\degree},~\SI{-30}{\degree},~\SI{30}{\degree},~\SI{70}{\degree}]^T$. Similar to the case of ideal autopilot, in this case the desired time of interception is chosen as \SI{30}{s}. The time-constant, $\tau$, of the autopilot dynamics is \SI{0.5}{s}. The free-will arbitrary time, $t_{f_{\ell}}$, within which the leader converges to the ideal time-to-go, is chosen as \SI{5}{s}. The free-will arbitrary time within which the followers are required to converge to the leader's time-to-go is chosen as $t_f = \SI{25}{s}$, while the time $t_1 = \SI{24}{s}$. The parameters $\eta_{\ell}$ and $\eta_{\ell_2}$ for the leader are 7 and 1.5, respectively, while those for the followers are $\eta_f = 15$ and $\eta_{f_2} = 4$. Due to the exponential terms present in \eqref{eq:second_order_control_leader} and \eqref{eq:second_order_follower_consensus_law}, large numerical values of time-to-go error for both leader and followers cause $u_{\ell}$ and $\bm{u_f}$ to rise to unacceptably high values. To alleviate this, we scaled both sides of \eqref{eq:second_order_tgo_dynamics_leader} and \eqref{eq:second_order_tgo_dynamics_follower} by a regularizing factor of 0.01 to avoid numerical errors in computations. The lateral acceleration command is assumed to be bounded within \SI{300}{m/s^2} for each interceptor.
\begin{figure*}[tb]
    \centering
    \begin{subfigure}[t]{0.32\linewidth}
    	\includegraphics[width=\textwidth] {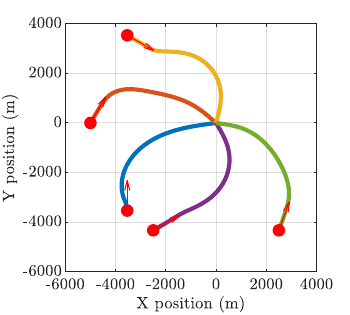}
    	\caption{Trajectory.}
    	\label{fig:double_integrator_xy}
    \end{subfigure}
    \begin{subfigure}[t]{0.32\linewidth}
    	\includegraphics[width=\textwidth] {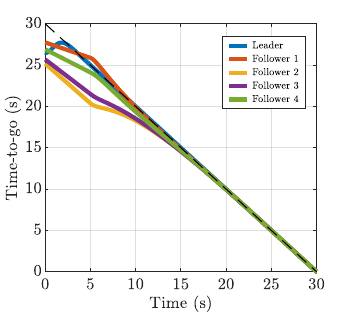}
    	\caption{Time-to-go.}
    	\label{fig:double_integrator_tgo}
    \end{subfigure}
    %
    %
    \begin{subfigure}[t]{0.32\linewidth}
    	\includegraphics[width=\textwidth] {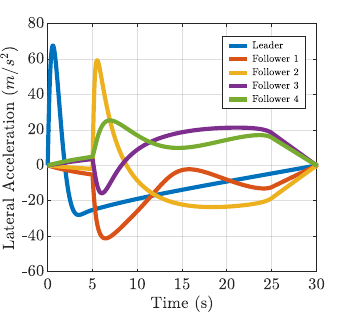}
    	\caption{Lateral acceleration.}
    	\label{fig:double_integrator_am}
    \end{subfigure}  
    %
    \caption{Cooperative guidance strategy for interceptors with first-order interceptor autopilot.}
    \label{fig:double_integrator}
\end{figure*}
Figure~\ref{fig:double_integrator_xy} shows the trajectories of the interceptors which start from their initial positions (marked by the red dots) and eventually intercept the target located at the origin, as before. Compared with Fig.~\ref{fig:single_integrator_xy}, the effect of the first-order autopilot is evident as the initial part of the leader's trajectory turns with a distinct lag. This is because, unlike for the ideal autopilot, the desired lateral acceleration is not achieved instantaneously in this case. Further, the follower interceptors start maneuvering to achieve consensus after the convergence of the leader's time-to-go to its desired value. 

It may be observed in Fig.~\ref{fig:double_integrator_tgo} that, using the proposed guidance strategy, all the interceptors intercept the target precisely at the desired impact time of \SI{30}{s} even though the initial time-to-go values for individual interceptors are much lower than \SI{30}{s}. In fact, the interceptors tend to veer away from the target initially to satisfy the impact time constraint. The time-to-go of the followers vary with their individual slopes until the consensus control $\bm{u_f}$ becomes active at $t_{f_{\ell}} = \SI{5}{s}$. Subsequently, the cooperative control law ensures that the time-to-go trajectories of the interceptors converge to the ideal expression, $t_{go} = 30-t$, within the prescribed \SI{25}{s}. The leader achieves the desired $\dot{t}_{go} = -1$ within the prespecified $t_{f_{\ell}} = \SI{5}{s}$, while the followers achieve $\dot{t}_{go} = -1$ within \SI{25}{s} as expected.  Figure~\ref{fig:double_integrator_am} shows the evolution of the lateral acceleration, $a_M$, in this case. 

Numerical simulations with different initial conditions and system parameters were presented in this section to assess the effectiveness of the proposed guidance schemes. The results indicate that even though the time for consensus can be chosen arbitrarily, the choice must be made judiciously considering the bounds on the control inputs.

\subsection{Simulations to demonstrate scalability of the algorithm}

To demonstrate the scalability of the salvo guidance algorithm, we have carried out simulations with double the number of follower interceptors and a fairly complex communication graph between them. The simulation conditions and results are discussed below.

\begin{figure*}[tb]
    \centering
    \begin{subfigure}[t]{0.48\linewidth}
	\centering
    	\includegraphics[width=0.9\textwidth]{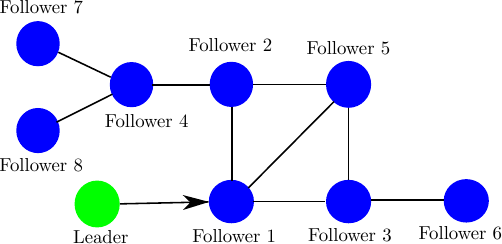}	
	\caption{Network graph with 1 leader and 8 followers.}
 	\label{fig:network_graph_8follower}
    \end{subfigure}
    \begin{subfigure}[t]{0.48\linewidth}
	\centering
    	\includegraphics[width=0.6\textwidth]{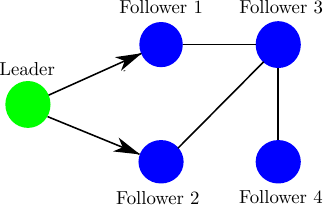}	
	\caption{Network graph with 1 leader and 4 followers.}
 	\label{fig:network_graph_comparison}
    \end{subfigure}
    \caption{Network graphs for simulations.}
    \label{fig:network_graph_8follower_comparison}
\end{figure*}

In this simulation scenario, there are 8 followers along with 1 leader. The communication graph among the leader and 8 followers are shown in Fig.~\ref{fig:network_graph_8follower} . All the interceptors are at an initial radial distance of \SI{5}{km} from the target with speed of \SI{200}{m/s}. The initial azimuth and heading angle of the leader are $45^{\circ}$ and $90^{\circ}$ respectively. The follower interceptors are located at initial azimuth angles of $0^{\circ}$, $-45^{\circ}$, $60^{\circ}$, $120^{\circ}$, $190^{\circ}$, $-110^{\circ}$, $90^{\circ}$, $160^{\circ}$, while their initial heading angles are $60^{\circ}$, $-30^{\circ}$, $30^{\circ}$, $70^{\circ}$, $130^{\circ}$, $240^{\circ}$, $20^{\circ}$, $100^{\circ}$. The desired time of interception is \SI{30}{s}, and the free-will arbitrary convergence times are $t_{f_l}=$\SI{5}{s} and  $t_f=$\SI{25}{s}. The parameters $\eta_l$ and $\eta_f$ are 3 and 24.7572, respectively.

\begin{figure*}[tb]
    \centering
    \begin{subfigure}[t]{0.32\linewidth}
    	\includegraphics[width=\textwidth]{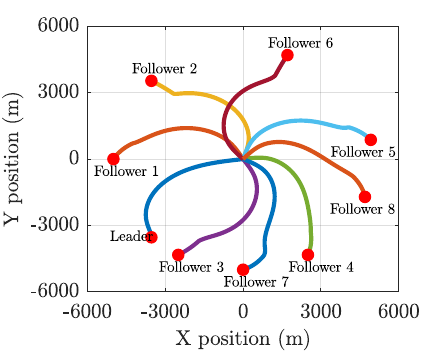}	
    	\caption{Trajectory.}
	    \label{fig:8follower_single_integrator_xy}
    \end{subfigure}
    \begin{subfigure}[t]{0.32\linewidth}
    	\includegraphics[width=\textwidth]{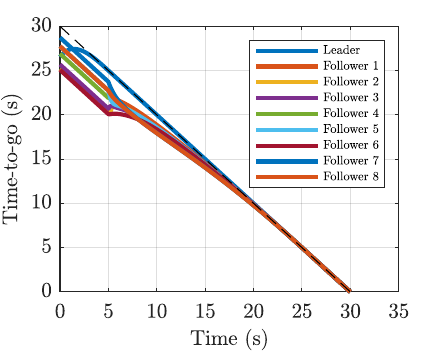}
    	\caption{Time-to-go.}
    	\label{fig:8follower_single_integrator_tgo}
    \end{subfigure}
    \begin{subfigure}[t]{0.32\linewidth}
    	\includegraphics[width=\textwidth]{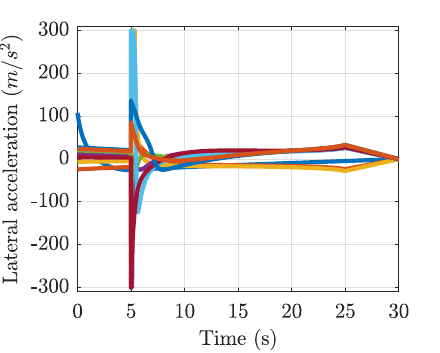}
    	\caption{Lateral acceleration command.}
    	\label{fig:8follower_single_integrator_am}
    \end{subfigure}
    \caption{Simulation results of cooperative guidance strategy for large number of interceptors.}
    \label{fig:8follower_single_integrator}
\end{figure*}

The simulation results are shown in Fig. \ref{fig:8follower_single_integrator}. As seen in Fig. \ref{fig:8follower_single_integrator_xy} illustrating the trajectories of the interceptors, successful interception of the target is achieved by each interceptor. Fig. \ref{fig:8follower_single_integrator_tgo} shows the evolution of time-to-go of the interceptors. The $t_{go}$ of the leader converges with the desired time-to-go curve within  $t_{f_l}=$\SI{5}{s}, and subsequently all the followers converge to the leader's time-to-go within the desired $t_f=$\SI{25}{s}. Thereafter, simultaneous interception of the target is achieved by all the interceptors at the desired impact time of \SI{30}{s}.

\subsection{Comparison with existing salvo guidance}

To the best of our knowledge, a salvo guidance design with free-will arbitrary time convergence in consensus is not present in the literature. To demonstrate the efficiency of our algorithm, its performance is compared with another salvo guidance algorithm described in Section 3.1.1 of \cite{Sinha2022}, which uses leader-follower communication framework in order to achieve interception with a stationary target. In order to have similar comparison of the interceptor dynamics, the simulations results with first-order autopilot is shown below, where the time-to-go dynamics has relative degree two with respect to the input command.

Same initial conditions are used in our simulations. There is a single leader (designated as $I_0$ in \cite{Sinha2022}) with 4 followers (designated as $I_1$ to $I_4$ in \cite{Sinha2022}) with the communication graph as shown in Fig. \ref{fig:network_graph_comparison}. We have considered bidirectional communication among the followers and unidirectional communication from the leader to the followers $I_1$ and $I_2$ as per Assumptions 2 and 3 in our paper. The initial radial distance of all the interceptors from the target is \SI{10}{km} and the speed of each interceptor is \SI{400}{m/s}. The initial azimuth angle and heading angle of the leader are $30^{\circ}$ and $0^{^\circ}$ respectively, while the followers are at the initial azimuth angles of $150^{\circ}$, $-30^{\circ}$, $210^{\circ}$, $45^{\circ}$ with the initial heading angles of  $170^{\circ}$,  $0^{\circ}$,  $250^{\circ}$,  $90^{\circ}$. The desired time of simultaneous interception with the stationary target is chosen as $T_d = $ \SI{30}{s}. 

\begin{figure*}[tb]
    \centering
    \begin{subfigure}[t]{0.32\linewidth}
    	\includegraphics[width=\textwidth]{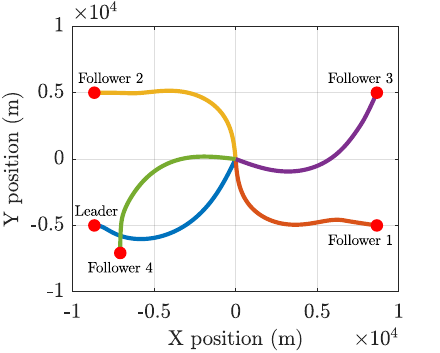}	
    	\caption{Trajectory.}
	    \label{fig:comparison_xy}
    \end{subfigure}
    \begin{subfigure}[t]{0.32\linewidth}
    	\includegraphics[width=\textwidth]{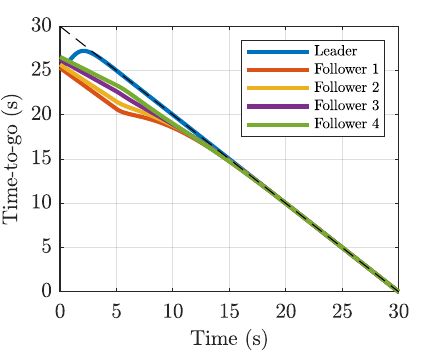}
    	\caption{Time-to-go.}
    	\label{fig:comparison_tgo}
    \end{subfigure}
    \begin{subfigure}[t]{0.32\linewidth}
    	\includegraphics[width=\textwidth]{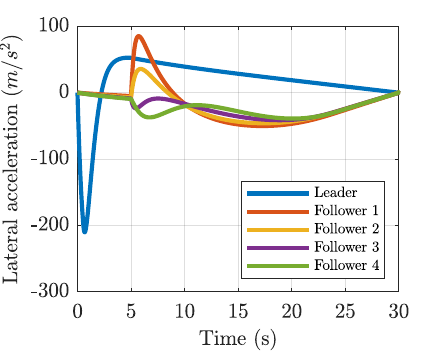}
    	\caption{Lateral acceleration command.}
    	\label{fig:comparison_am}
    \end{subfigure}
    \caption{Simulation results with initial conditions given in \cite{Sinha2022}.}
    \label{fig:comparison}
\end{figure*}

As seen in Fig. \ref{fig:comparison_xy} all the interceptors achieve target interception at the desired time $T_d$ of \SI{30}{s}. The initial time-to-go of the interceptors are \SI{25.68}{s}, \SI{25.30}{s}, \SI{25.68}{s}, \SI{26.22}{s} and \SI{26.54}{s}, which are very close to those reported in \cite{Sinha2022}, which proves that the small angle approximation in time-to-go in our paper is not too detrimental to its performance. Fig. \ref{fig:comparison_tgo} shows that the leader's time-to-go converges on the desired time-to-go curve within \SI{5}{s} and the followers' time-to-go converge on the leader's time-to-go within \SI{25}{s} as chosen in our algorithm. Similar to \cite{Sinha2022}, the lateral acceleration at the terminal time, as shown in Fig. \ref{fig:comparison_am} is zero, which is a desirable feature. However, unlike \cite{Sinha2022}, the variation of lateral acceleration in our simulations are significantly smoother. This is definitely advantageous with respect to structural flexibility design of the interceptors.

\section{Conclusion} 
\label{sec:conclusion}

A cooperative guidance algorithm was developed in this paper for a salvo attack against a stationary target at a desired impact time by multiple interceptors that share information with each other over a leader-follower communication framework. The guidance strategies were designed using time-to-go estimates for interceptors with two types of dynamics -- ideal autopilot and first-order autopilot. Stability analyses established the convergence of the leader's time-to-go to the ideal time-to-go trajectory and that of the followers to the leader's time-to-go within a pre-specified time that the designer can choose freely and is not dependent on the initial conditions or other design parameters. Numerical simulations with different initial conditions and system parameters were performed to assess the effectiveness of the proposed guidance scheme. 

The consensus control command tends to be large when the difference in times-to-go between a pair of the followers is high. This is due to the presence of the exponential term present in the consensus control law. When this exponential term is high enough such that it increases beyond the practical limit of the interceptor's control, the command saturates. This saturation in control command will affect pre-specified time convergence only if there is insufficient time for the interceptors' maneuvers. Therefore, this can be mitigated by choosing the desired interception time, the free-will arbitrary pre-specified times and the gains judiciously, considering the bounds on the control inputs, interceptor's speed, initial position, maximum achievable lateral acceleration, and other physical constraints. It is evident from the simulations that even though command saturation has occurred in most of the cases, the interceptors are able to achieve salvo interception of the target in all the cases. Future research shall focus on developing control algorithms that consider the bounds of control commands in the formulation itself. 

Developing cooperative salvo guidance strategies using consensus over general directed communication topologies with maneuvering targets shall be pursued in the future. Development of the salvo algorithm taking into account delay, degradation or failures in communication among the interceptors might be another possible direction of future research.

Choice of the leader should be made judiciously based on the initial time-to-go of the interceptors. It is well known that any leader-follower based consensus algorithms suffers from the inherent disadvantage that failure of the leader results in failure of all its followers as well. Design of the salvo algorithm considering leader failure is an interesting topic for future research.

\bibliographystyle{IEEEtran}
\bibliography{reference_tase_2023}

\begin{thebibliography}{10}
\providecommand{\url}[1]{#1}
\csname url@samestyle\endcsname
\providecommand{\newblock}{\relax}
\providecommand{\bibinfo}[2]{#2}
\providecommand{\BIBentrySTDinterwordspacing}{\spaceskip=0pt\relax}
\providecommand{\BIBentryALTinterwordstretchfactor}{4}
\providecommand{\BIBentryALTinterwordspacing}{\spaceskip=\fontdimen2\font plus
\BIBentryALTinterwordstretchfactor\fontdimen3\font minus
  \fontdimen4\font\relax}
\providecommand{\BIBforeignlanguage}[2]{{%
\expandafter\ifx\csname l@#1\endcsname\relax
\typeout{** WARNING: IEEEtran.bst: No hyphenation pattern has been}%
\typeout{** loaded for the language `#1'. Using the pattern for}%
\typeout{** the default language instead.}%
\else
\language=\csname l@#1\endcsname
\fi
#2}}
\providecommand{\BIBdecl}{\relax}
\BIBdecl

\bibitem{Zhang2015}
Y.~Zhang, X.~Wang, and H.~Wu, ``A distributed cooperative guidance law for
  salvo attack of multiple anti-ship missiles,'' \emph{Chinese Journal of
  Aeronautics}, vol.~28, no.~5, pp. 1438--1450, Oct. 2015.

\bibitem{Song2015}
L.~Song, Y.~Zhang, D.~Huang, and S.~Fu, ``Cooperative simultaneous attack of
  multi-missiles under unreliable and noisy communication network: {{A}}
  consensus scheme of impact time,'' \emph{Aerospace Science and Technology},
  vol.~47, pp. 31--41, Dec. 2015.

\bibitem{Zhu2017}
Q.~Zhu, X.~Wang, and Q.~Lin, ``Consensus-based impact-time-control guidance law
  for cooperative attack of multiple missiles,'' \emph{Kybernetika}, vol.~53,
  no.~4, pp. 563--577, 2017.

\bibitem{He2018a}
S.~He, W.~Wang, D.~Lin, and H.~Lei, ``Consensus-{{Based Two-Stage Salvo Attack
  Guidance}},'' \emph{IEEE Transactions on Aerospace and Electronic Systems},
  vol.~54, no.~3, pp. 1555--1566, Jun. 2018.

\bibitem{He2018}
S.~He, M.~Kim, T.~Song, and D.~Lin, ``Three-dimensional salvo attack guidance
  considering communication delay,'' \emph{Aerospace Science and Technology},
  vol.~73, pp. 1--9, Feb. 2018.

\bibitem{Ai2019}
X.~Ai, L.~Wang, J.~Yu, and Y.~Shen, ``Field-of-view constrained two-stage
  guidance law design for three-dimensional salvo attack of multiple missiles
  via an optimal control approach,'' \emph{Aerospace Science and Technology},
  vol.~85, pp. 334--346, Feb. 2019.

\bibitem{Jeon2010}
I.-S. Jeon, J.-I. Lee, and M.-J. Tahk, ``Homing {{Guidance Law}} for
  {{Cooperative Attack}} of {{Multiple Missiles}},'' \emph{Journal of Guidance,
  Control, and Dynamics}, vol.~33, no.~1, pp. 275--280, 2010.

\bibitem{Hou2015}
D.~Hou, Q.~Wang, X.~Sun, and C.~Dong, ``Finite-time cooperative guidance laws
  for multiple missiles with acceleration saturation constraints,'' \emph{IET
  Control Theory \& Applications}, vol.~9, no.~10, pp. 1525--1535, 2015.

\bibitem{Li2016}
B.~Li, D.~Lin, and H.~Wang, ``Finite time convergence cooperative guidance law
  based on graph theory,'' \emph{Optik}, vol. 127, no.~21, pp.
  10\,180--10\,188, Nov. 2016.

\bibitem{Kumar2020}
S.~R. Kumar and D.~Mukherjee, ``Cooperative {{Salvo Guidance Using Finite-Time
  Consensus}} over {{Directed Cycles}},'' \emph{IEEE Transactions on Aerospace
  and Electronic Systems}, vol.~56, no.~2, pp. 1504--1514, 2020.

\bibitem{Li2018a}
G.~Li, Y.~Wu, and P.~Xu, ``Adaptive fault-tolerant cooperative guidance law for
  simultaneous arrival,'' \emph{Aerospace Science and Technology}, vol. 82--83,
  pp. 243--251, Nov. 2018.

\bibitem{Kumar2015}
S.~R. Kumar and D.~Ghose, ``Impact time guidance for large heading errors using
  sliding mode control,'' \emph{IEEE Transactions on Aerospace and Electronic
  Systems}, vol.~51, no.~4, pp. 3123--3138, Oct. 2015.

\bibitem{Kim2019}
H.-G. Kim, D.~Cho, and H.~J. Kim, ``Sliding {{Mode Guidance Law}} for {{Impact
  Time Control Without Explicit Time-to-Go Estimation}},'' \emph{IEEE
  Transactions on Aerospace and Electronic Systems}, vol.~55, no.~1, pp.
  236--250, Feb. 2019.

\bibitem{Chen2018}
X.~Chen and J.~Wang, ``Nonsingular {{Sliding-Mode Control}} for {{Field-of-View
  Constrained Impact Time Guidance}},'' \emph{Journal of Guidance, Control, and
  Dynamics}, vol.~41, no.~5, pp. 1214--1222, May 2018.

\bibitem{Hu2019}
Q.~Hu, T.~Han, and M.~Xin, ``Sliding-{{Mode Impact Time Guidance Law Design}}
  for {{Various Target Motions}},'' \emph{Journal of Guidance, Control, and
  Dynamics}, vol.~42, no.~1, pp. 136--148, Jan. 2019.

\bibitem{Jeon2006}
I.-S. Jeon, J.-I. Lee, and M.-J. Tahk, ``Impact-time-control guidance law for
  anti-ship missiles,'' \emph{IEEE Transactions on Control Systems Technology},
  vol.~14, no.~2, pp. 260--266, Mar. 2006.

\bibitem{Liu2017b}
X.~Liu, Z.~Shen, and P.~Lu, ``Closed-{{Loop Optimization}} of {{Guidance Gain}}
  for {{Constrained Impact}},'' \emph{Journal of Guidance, Control, and
  Dynamics}, vol.~40, no.~2, pp. 453--460, Feb. 2017.

\bibitem{Lu2006}
P.~Lu, D.~B. Doman, and J.~D. Schierman, ``Adaptive {{Terminal Guidance}} for
  {{Hypervelocity Impact}} in {{Specified Direction}},'' \emph{Journal of
  Guidance, Control, and Dynamics}, vol.~29, no.~2, pp. 269--278, Mar. 2006.

\bibitem{Yang2023}
X.~Yang, Y.~Zhang, and S.~Song, ``Two-{{Stage Cooperative Guidance Strategy}}
  with {{Impact-Angle}} and {{Field-of-View Constraints}},'' \emph{Journal of
  Guidance, Control, and Dynamics}, vol.~46, no.~3, pp. 590--599, Mar. 2023.

\bibitem{Bhat2000}
S.~P. Bhat and D.~S. Bernstein, ``Finite-{{Time Stability}} of {{Continuous
  Autonomous Systems}},'' \emph{SIAM J. Control Optim.}, vol.~38, no.~3, pp.
  751--766, Jan. 2000.

\bibitem{Zhou2009}
D.~Zhou, S.~Sun, and K.~L. Teo, ``Guidance {{Laws}} with {{Finite Time
  Convergence}},'' \emph{Journal of Guidance, Control, and Dynamics}, vol.~32,
  no.~6, pp. 1838--1846, Nov. 2009.

\bibitem{Polyakov2012}
A.~Polyakov, ``Nonlinear feedback design for fixed-time stabilization of linear
  control systems,'' \emph{IEEE Transactions on Automatic Control}, vol.~57,
  no.~8, pp. 2106--2110, Aug. 2012.

\bibitem{Zuo2018}
Z.~Zuo, Q.~L. Han, B.~Ning, X.~Ge, and X.~M. Zhang, ``An overview of recent
  advances in fixed-time cooperative control of multiagent systems,''
  \emph{IEEE Transactions on Industrial Informatics}, vol.~14, no.~6, pp.
  2322--2334, 2018.

\bibitem{Wang2022a}
C.~Wang, W.~Dong, J.~Wang, and M.~Xin, ``Impact-{{Angle-Constrained Cooperative
  Guidance}} for {{Salvo Attack}},'' \emph{Journal of Guidance, Control, and
  Dynamics}, vol.~45, no.~4, pp. 684--703, Apr. 2022.

\bibitem{Pal2020a}
A.~K. Pal, S.~Kamal, S.~K.~S. Nagar, B.~Bandyopadhyay, and L.~Fridman, ``Design
  of controllers with arbitrary convergence time,'' \emph{Automatica}, vol.
  112, p. 108710, 2020.

\bibitem{Tran2022}
Q.~V. Tran, M.~H. Trinh, N.~H. Nguyen, and H.-S. Ahn, ``Free-will arbitrary
  time consensus protocols with diffusive coupling,'' \emph{International
  Journal of Robust and Nonlinear Control}, vol.~32, no.~15, pp. 8711--8731,
  2022.

\bibitem{Pal2022}
A.~K. Pal, S.~Kamal, X.~Yu, S.~K. Nagar, and X.~Xiong, ``Free-will arbitrary
  time consensus for multiagent systems,'' \emph{IEEE Transactions on
  Cybernetics}, vol.~52, no.~6, pp. 4636--4646, Jun. 2022.

\bibitem{pal2023acc}
R.~S. Pal, S.~R. Kumar, and D.~Mukherjee, ``Free will arbitrary time
  consensus-based cooperative salvo guidance over leader-follower network,'' in
  \emph{American Control Conference}.\hskip 1em plus 0.5em minus 0.4em\relax
  (To~ appear~ online): IEEE, 2023, pp. 4936--4941.

\bibitem{Godsil2001}
C.~Godsil and G.~Royle, \emph{Algebraic Graph Theory}, ser. Graduate {{Texts}}
  in {{Mathematics}}.\hskip 1em plus 0.5em minus 0.4em\relax {New York, NY}:
  {Springer}, 2001, vol. 207.

\bibitem{Sinha2022}
A.~Sinha, S.~R. Kumar, and D.~Mukherjee, ``Cooperative integrated guidance and
  control design for simultaneous interception,'' \emph{Aerospace Science and
  Technology}, vol. 120, p. 107262, Jan. 2022.

\end{thebibliography}

 \begin{IEEEbiography}[{\includegraphics[width=25mm,height=32mm,clip,keepaspectratio]{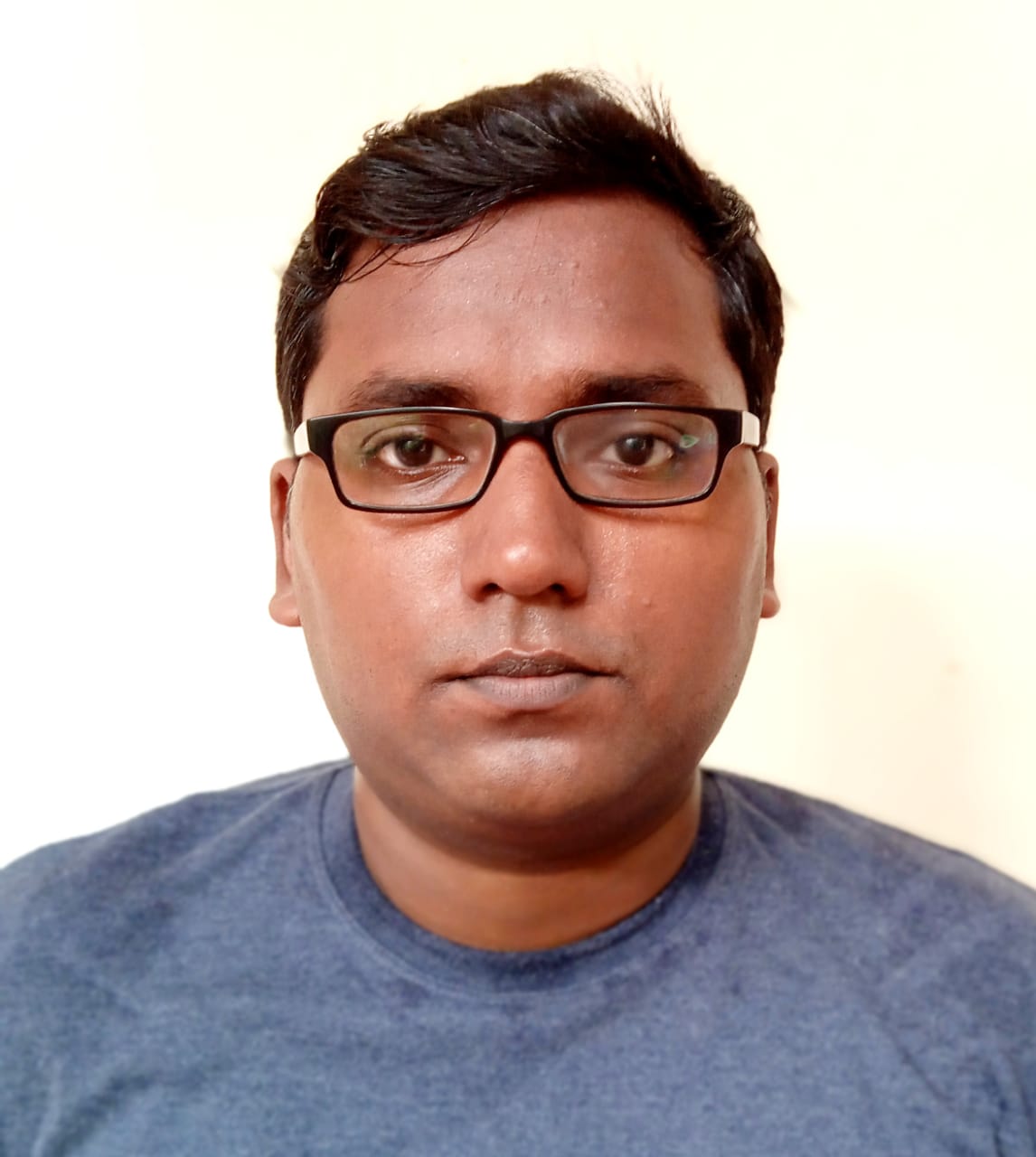}}]
 	{\bf Rajib~Shekhar~Pal} is a Research Scholar of Aerospace Engineering at the Indian Institute of Technology Bombay. He received his B.Tech. (2013) degree in Aerospace Engineering from Indian Institute of Space Science and Technology, India. His research interests include topics related to the guidance of aerospace vehicles, path planning, multi-agent systems, and cooperative control.
 \end{IEEEbiography}

 \begin{IEEEbiography}[{\includegraphics[width=25mm,height=32mm,clip,keepaspectratio]{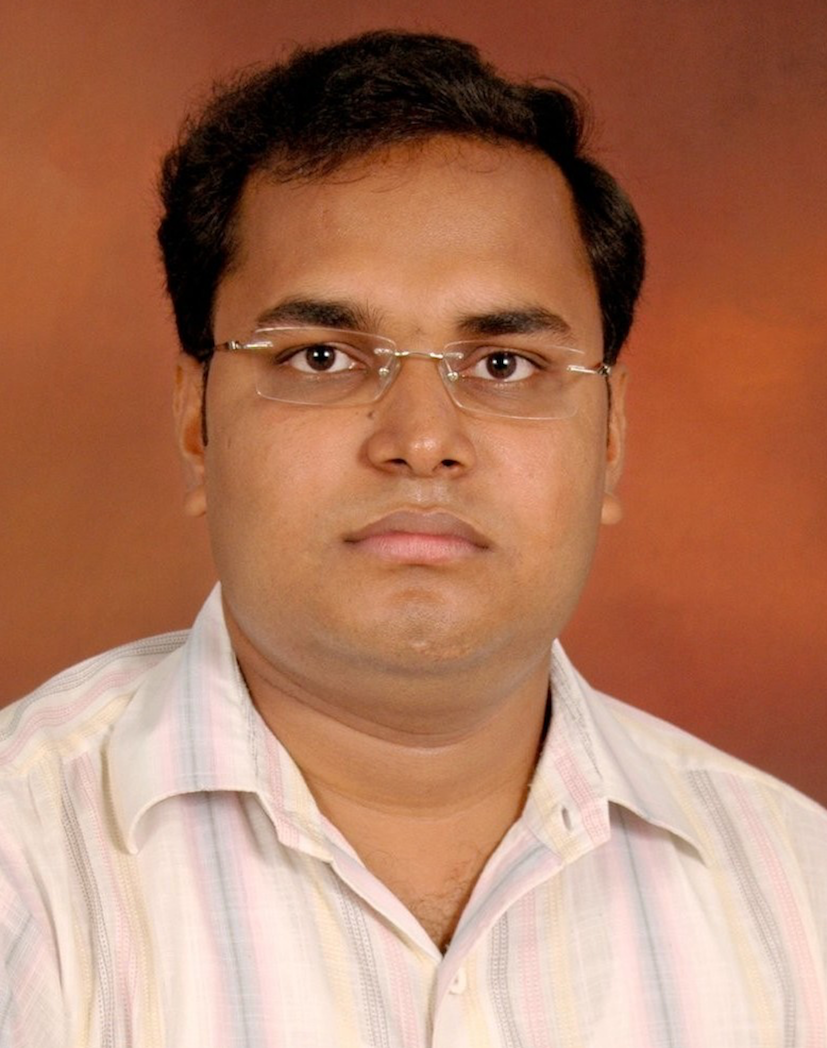}}]
 	{\bf Shashi~Ranjan~Kumar} is an Associate Professor of Aerospace Engineering at the Indian Institute of Technology Bombay. He received his B.Sc. (Engg.) (2008) degree in Electronics and Communication Engineering from Muzaffarpur Institute of Technology, India. He completed his M.E. (2010) and Ph.D. (2015), both in Aerospace Engineering, from the Indian Institute of Science, Bangalore. From 2015-2017, he was a post-doctoral fellow at the Faculty of Aerospace Engineering, Technion $-$ Israel Institute of Technology. His research interests include topics related to the guidance of aerospace vehicles, path planning, multi-agent systems, and cooperative control.
 \end{IEEEbiography}
 
 \begin{IEEEbiography}[{\includegraphics[width=25mm,height=32mm,clip,keepaspectratio]{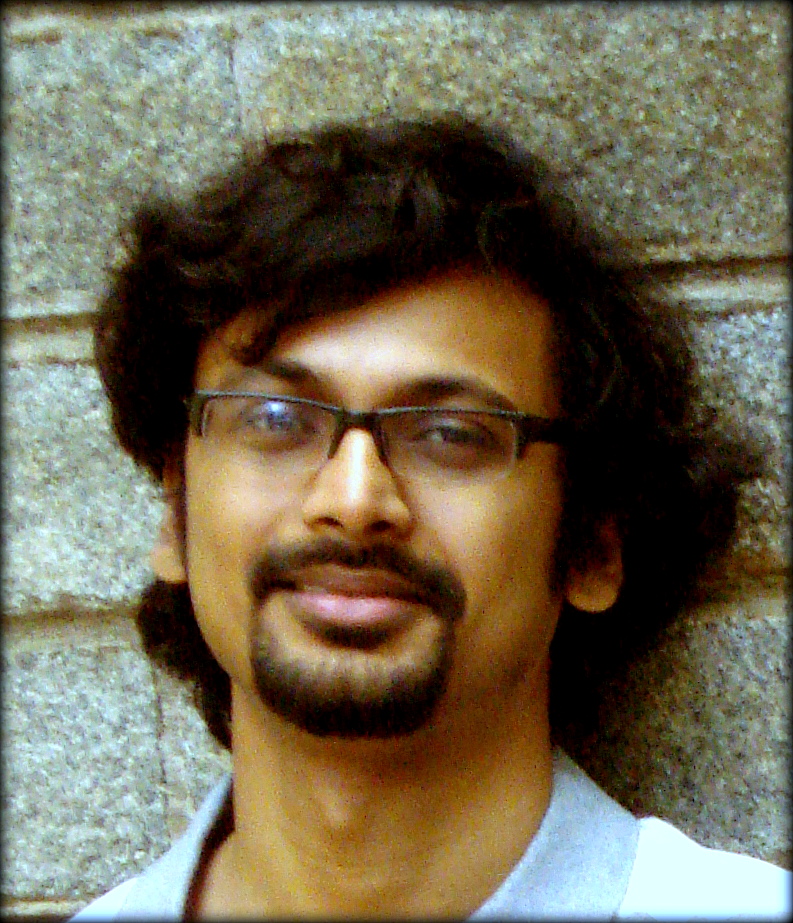}}]
 	{\bf Dwaipayan~Mukherjee} is an Associate Professor of Electrical Engineering at the Indian Institute of Technology Bombay. He received his B.E. (2007) from Jadavpur University, Kolkata, in Electrical Engineering and M.Tech. (2009) in Control Systems Engineering from the Indian Institute of Technology Kharagpur. In 2014, he completed his Ph.D. in Aerospace Engineering, from the Indian Institute of Science, Bangalore. From 2015-2017, he was a post-doctoral fellow at the Faculty of Aerospace Engineering, Technion $-$ Israel Institute of Technology. His research interests include multi-agent systems, cooperative control, and control theory.
 \end{IEEEbiography}

\vfill

\end{document}